\documentstyle[11pt]{article}
\newcounter{subequation}[equation]

\makeatletter 

\def\citen#1{%
\edef\@tempa{\@ignspaftercomma,#1, \@end, }
\edef\@tempa{\expandafter\@ignendcommas\@tempa\@end}%
\if@filesw \immediate \write \@auxout {\string \citation {\@tempa}}\fi
\@tempcntb\m@ne \let\@h@ld\relax \def\@citea{}%
\@for \@citeb:=\@tempa\do {\@cmpresscites}%
\@h@ld}
\def\@ignspaftercomma#1, {\ifx\@end#1\@empty\else
   #1,\expandafter\@ignspaftercomma\fi}
\def\@ignendcommas,#1,\@end{#1}
\def\@cmpresscites{%
 \expandafter\let \expandafter\@B@citeB \csname b@\@citeb \endcsname
 \ifx\@B@citeB\relax 
    \@h@ld\@citea\@tempcntb\m@ne{\bf ?}%
    \@warning {Citation `\@citeb ' on page \thepage \space undefined}%
 \else
    \@tempcnta\@tempcntb \advance\@tempcnta\@ne
    \setbox\z@\hbox\bgroup 
    \ifnum0<0\@B@citeB \relax
       \egroup \@tempcntb\@B@citeB \relax
       \else \egroup \@tempcntb\m@ne \fi
    \ifnum\@tempcnta=\@tempcntb 
       \ifx\@h@ld\relax 
          \edef \@h@ld{\@citea\@B@citeB }%
       \else 
          \edef\@h@ld{\hbox{--}\penalty\@highpenalty
            \@B@citeB }%
       \fi
    \else   
       \@h@ld\@citea\@B@citeB
       \let\@h@ld\relax
 \fi\fi%
 \def\@citea{,\penalty\@highpenalty\hskip.13em plus.1em minus.1em}%
}
\def\@citex[#1]#2{\@cite{\citen{#2}}{#1}}%
\def\@cite#1#2{\leavevmode\unskip
  \ifnum\lastpenalty=\z@\penalty\@highpenalty\fi
  \ [{\multiply\@highpenalty 3 #1
      \if@tempswa,\penalty\@highpenalty\ #2\fi 
    }]\spacefactor\@m}
\def\thesubequation{\theequation\@alph\c@subequation}
\def\@subeqnnum{{\rm (\thesubequation)}}
\def\slabel#1{\@bsphack\if@filesw {\let\thepage\relax
   \xdef\@gtempa{\write\@auxout{\string
      \newlabel{#1}{{\thesubequation}{\thepage}}}}}\@gtempa
   \if@nobreak \ifvmode\nobreak\fi\fi\fi\@esphack}
\def\subeqnarray{\stepcounter{equation}
\let\@currentlabel=\theequation\global\c@subequation\@ne
\global\@eqnswtrue
\global\@eqcnt\z@\tabskip\@centering\let\\=\@subeqncr
$$\halign to \displaywidth\bgroup\@eqnsel\hskip\@centering
  $\displaystyle\tabskip\z@{##}$&\global\@eqcnt\@ne
  \hskip 2\arraycolsep \hfil${##}$\hfil
  &\global\@eqcnt\tw@ \hskip 2\arraycolsep
  $\displaystyle\tabskip\z@{##}$\hfil
   \tabskip\@centering&\llap{##}\tabskip\z@\cr}
\def\endsubeqnarray{\@@subeqncr\egroup
                     $$\global\@ignoretrue}
\def\@subeqncr{{\ifnum0=`}\fi\@ifstar{\global\@eqpen\@M
    \@ysubeqncr}{\global\@eqpen\interdisplaylinepenalty \@ysubeqncr}}
\def\@ysubeqncr{\@ifnextchar [{\@xsubeqncr}{\@xsubeqncr[\z@]}}
\def\@xsubeqncr[#1]{\ifnum0=`{\fi}\@@subeqncr
   \noalign{\penalty\@eqpen\vskip\jot\vskip #1\relax}}
\def\@@subeqncr{\let\@tempa\relax
    \ifcase\@eqcnt \def\@tempa{& & &}\or \def\@tempa{& &}
      \else \def\@tempa{&}\fi
     \@tempa \if@eqnsw\@subeqnnum\refstepcounter{subequation}\fi
     \global\@eqnswtrue\global\@eqcnt\z@\cr}
\let\@ssubeqncr=\@subeqncr
\@namedef{subeqnarray*}{\def\@subeqncr{\nonumber\@ssubeqncr}\subeqnarray}
\@namedef{endsubeqnarray*}{\global\advance\c@equation\m@ne%
                           \nonumber\endsubeqnarray}

\@addtoreset{equation}{section}
\renewcommand{\theequation}{\thesection.\arabic{equation}}

\makeatother

\def\dalemb#1#2{{\vbox{\hrule height .#2pt
        \hbox{\vrule width.#2pt height#1pt \kern#1pt
                \vrule width.#2pt}
        \hrule height.#2pt}}}
\def\square{\mathord{\dalemb{6.8}{7}\hbox{\hskip1pt}}}

\let\a=\alpha \let\b=\beta   \let\e=\epsilon
  \let\q=\theta \let\iota=\iota 
  \let\n=\nu

 \let\Pi=\Pi \let\Sigma=\Sigma

\def\nn{\nonumber} \def\bd{\begin{document}} \def\ed{\end{document}}
\def\ds{\documentstyle} \let\fr=\frac \let\bl=\bigl \let\br=\bigr
\let\Br=\Bigr \let\Bl=\Bigl 
\let\bm=\bibitem
\let\na=\nabla
\let\pa=\partial \let\ov=\overline
\def\ie{{\it i.e.\ }} 
\newcommand{\be}{\begin{equation}} 
\newcommand{\ee}{\end{equation}}
\def\ba{\begin{array}}
\def\ea{\end{array}}
\def\ft#1#2{{\textstyle{{\scriptstyle #1}\over {\scriptstyle #2}}}}
\def\fft#1#2{{#1 \over #2}}
\def\del{\partial}
\def\st#1{{\scriptstyle #1}}
\def\sst#1{{\scriptscriptstyle #1}}
\def\oneone{\rlap 1\mkern4mu{\rm l}}
\def\e7{E_{7(+7)}}
\def\td{\tilde}
\def\wtd{\widetilde}
\def\im{{\rm i}}
\def\bog{Bogomol'nyi\ }
\def\q{{\tilde q}}
\def\hast{{\hat\ast}}
\def\0{{\sst{(0)}}}
\def\1{{\sst{(1)}}}
\def\2{{\sst{(2)}}}
\def\3{{\sst{(3)}}}
\def\4{{\sst{(4)}}}
\def\5{{\sst{(5)}}}
\def\6{{\sst{(6)}}}
\def\7{{\sst{(7)}}}
\def\8{{\sst{(8)}}}
\def\n{{\sst{(n)}}}
\def\tV{{\wtd V}}

\newcommand{\ho}[1]{$\, ^{#1}$}
\newcommand{\hoch}[1]{$\, ^{#1}$}
\newcommand{\bea}{\begin{eqnarray}} 
\newcommand{\eea}{\end{eqnarray}}
\newcommand{\bsea}{\begin{subeqnarray}}
\newcommand{\esea}{\end{subeqnarray}}
\newcommand{\ra}{\rightarrow}
\newcommand{\lra}{\longrightarrow}
\newcommand{\Lra}{\Leftrightarrow}
\newcommand{\ap}{\alpha^\prime}
\newcommand{\bp}{\tilde \beta^\prime}
\newcommand{\tr}{{\rm tr} }
\newcommand{\Tr}{{\rm Tr} } 
\newcommand{\NP}{Nucl. Phys. }
\newcommand{\tamphys}{\it Center for Theoretical Physics,
Texas A\&M University, College Station, Texas 77843}
\newcommand{\ens}{\it Laboratoire de Physique Th\'eorique de l'\'Ecole
Normale Sup\'erieure\hoch{4,5}\\
24 Rue Lhomond - 75231 Paris CEDEX 05}
\newcommand{\sissa}{\it SISSA, Via Beirut No. 2-4, 34013 Trieste, 
Italy\hoch{4}}

\newcommand{\auth}{M.S. Bremer\hoch{\star\sharp\,1}, M.J.
Duff\hoch{\dagger\,2},  H. L\"u\hoch{\ddagger}, 
C.N. Pope\hoch{\dagger\S\,3} and K.S. Stelle\hoch{\star\sharp}}

\begin{document}
\begin{flushright}
\hfill{CERN-TH/98-194, \ \ CTP TAMU-27/98,\ \  
Imperial/TP/97-98/57}\\
\hfill{ LPTENS-98/19, \ \ SISSARef.\ 75/98/EP}\\
\hfill{\bf hep-th/9807051}\\
\hfill{July 1998}\\
\end{flushright}


\begin{center}
{\Large {\bf Instanton Cosmology and Domain Walls from\linebreak 
M-theory and String Theory}}

\vspace{15pt}

\auth

\vspace{10pt}
{\hoch{\star} \it The Blackett Laboratory, Imperial College\hoch{4}\\
Prince Consort Road, London SW7 2BZ, UK}

\vspace{7pt}
{\hoch{\sharp} \it TH Division, 
CERN, CH-1211 Geneva 23, Switzerland}

\vspace{7pt}
{\hoch{\dagger}\tamphys}

\vspace{7pt}
{\hoch{\ddagger}\ens}

\vspace{7pt}
{\hoch{\S}\sissa}

\vspace{20pt}

\underline{ABSTRACT}
\end{center}

    The recent proposal by Hawking and Turok for obtaining an open
inflationary universe from singular instantons makes use of low-energy
effective Lagrangians describing gravity coupled to scalars and
non-propagating antisymmetric tensors.  In this paper we derive some
exact results for Lagrangians of this type, obtained from spherical
compactifications of M-theory and string theory.  In the case of the
$S^7$ compactification of M-theory, we give a detailed discussion of
the cosmological solutions. We also show that the lower-dimensional
Lagrangians admit domain-wall solutions, which preserve one half of
the supersymmetry, and which approach AdS spacetimes near their
horizons.

{\vfill\leftline{}\vfill
\vskip  10pt
\footnoterule
{\footnotesize 
     \hoch{1} Research supported in part by the EC under TMR
contract ERBFMBI-CT97-2344. \vskip -12pt} \vskip 14pt
{\footnotesize
        \hoch{2} Research supported in part by NSF Grant PHY-9722090.
\vskip  -12pt} \vskip   14pt
{\footnotesize
        \hoch{3}        Research supported in part by DOE 
grant DE-FG03-95ER40917. \vskip -12pt}  \vskip  14pt
{\footnotesize
        \hoch{4} Research supported in part by the EC under TMR
contract ERBFMRX-CT96-0045. \vskip -12pt} \vskip 14pt
{\footnotesize
        \hoch{5} Unit\'e Propre du Centre National de la Recherche
Scientifique, associ\'ee \`a l'\'Ecole Normale Sup\'erieure
\vskip -12pt} \vskip 10pt
{\footnotesize \hoch{\phantom{3}} et \`a l'Universit\'e de Paris-Sud.
\vskip -12pt} \vskip 10pt}

\pagebreak
\setcounter{page}{1}

\tableofcontents
\addtocontents{toc}{\protect\setcounter{tocdepth}{2}}
\newpage

\section{Introduction\label{sec:intro}}

    It has recently been proposed that a Euclidean-signature instanton
solution can describe the creation of an open inflationary universe
\cite{ht1}.  The proposal has sparked a lively debate, in which
various aspects of the assumptions and the conclusions have been
discussed
\cite{linde1,ht2,ht3,unruh,boulin,garr1,vil1,vil2,garr2,gonz,smc}.

    One of the hopes of string theory or M-theory is that it will be
able to supply definitive answers to questions that until now have
been matters for conjecture or choice.  Indeed in \cite{ht3} the form
of the potential governing the evolution of an inflationary instanton
solution was drawn from M-theory, or, more specifically, from the
eleven-dimensional supergravity that is presumed to describe its
low-energy limit, by compactifying the theory on a 7-sphere.  This
makes use of the observation that a 4-form field strength can give
rise to a cosmological constant
\cite{duffvan,aurelia,fr,hawking,brown,duff,duncan}. Our purpose here
is not to step out into the fray in the ongoing cosmological debate,
but rather to study in more detail the lessons that can be learned by
taking M-theory or string theory as a starting-point for the
discussion.  We shall consider various examples in which
higher-dimensional supergravities are dimensionally reduced on
spheres, leading to theories in lower dimensions that include scalar
fields with the kind of potentials that are commonly encountered in
the inflationary models.  Of course, the complete analysis of the
dimensionally-reduced theories is extremely complicated, even at the
linearised level where one is simply concerned with extracting the
mass spectrum.  In fact for the purposes of studying cosmological
solutions it is not sufficient to know just the linearised results,
since the scalar fields are liable to become large at some stage in
the evolution of the system.  Thus it is of more interest to know the
exact form of the lower-dimensional Lagrangian for some subset of the
fields that includes the ones participating in the cosmological
solution.  In particular, if one wishes to retain comparability
between the different dimensional theories, it is important that the
truncation to the subset of fields in the lower-dimensional theory be
a {\it consistent} one, in the sense that solutions of the resulting
lower-dimensional equations of motion should also give solutions of
the original higher-dimensional ones.

    The easiest way of achieving a consistent truncation in any system
of equations is to retain the totality of fields that are singlets
under some symmetry group, while setting to zero all fields that are
non-singlets \cite{dp}.  By this means the danger inherent in any
``generic'' truncation, namely that non-linear terms can imply that
the retained fields act as sources for the fields that were set to
zero, is avoided.\footnote{In the context of a supergravity theory,
one also encounters many fields transforming nontrivially under
internal symmetry groups. Although such fields will not play a r\^ole
in the specific solutions that will occupy us in the present paper, it
is appropriate to note that truncation to the full supergravity
multiplet is also consistent \cite{dnp,zilch}.}
We shall consider a number of examples, based on sphere
compactifications of higher-dimensional supergravities.  The simplest
cases are where the theory is compactified on a ``round'' n-sphere
with $SO(n+1)$ isometry group, and the dimensionally-reduced theory is
truncated to the $SO(n+1)$ singlets.  In particular, the
Einstein-Hilbert sector of the higher-dimensional theory will give
rise just to the metric, and a ``breathing-mode'' scalar in the lower
dimension.  Slightly more complicated examples can be found in cases
where the dimension $n$ of the sphere $S^n$ is odd, $n=2m+1$.  In such
cases, we may truncate to the larger subsector of all singlets under
the $SU(m+1)\times U(1)$ subgroup of $SO(2m+2)$.  This is the isometry
group of the one-parameter family of homogeneous metrics on the
``squashed'' $(2m+1)$-sphere, which can be described as a $U(1)$
bundle over $CP^m$.  (The squashing parameter corresponds to the
freedom to scale the length of the $U(1)$ fibres, without affecting
the isometry group.  For one particular value of this parameter,
corresponding to the ``round'' sphere, the isometry group undergoes an
enlargement from $SU(m+1)\times U(1)$ to $SO(2m+2)$.)  Thus we may
construct consistently-truncated lower-dimensional theories that
include the homogeneous squashing mode as well as the breathing mode
of the sphere.

   In this paper, we shall construct consistently-truncated theories
that result from spherical reductions in a number of cases, including
the round and squashed $S^7$ reductions of $D=11$ supergravity, the
$S^4$ reduction of $D=11$ supergravity, and the round and squashed
$S^5$ reductions of type IIB supergravity. Owing to the presence of
cosmological potentials in these dimensionally reduced theories, there
is a salient class of supersymmetric solutions containing $(D-2)$
branes, \ie domain walls. Unlike the domain walls occurring in massive
supergravities \cite{masswalls,lpdomain}, these domain walls have
regular dilatons on their horizons.  Thus, they fit into the class of
solitons interpolating between different vacua of the dimensionally
reduced theory, specifically, between anti-de Sitter space and flat
space. Accordingly, they also display supersymmetry enhancement at
their horizons. We shall show that these domain-wall solutions can be
interpreted {\it via} dimensional oxidation on spheres, in terms of
the familiar fundamental brane solutions of $D=11$ and $D=10$ type IIB
supergravities. Next, we proceed to consider instanton solutions
obtained in two variants of Euclidean supergravities: those obtained
by Wick rotation and those obtained by dimensional reduction on
anti-de Sitter spaces, a Minkowski-signature variant of the spherical
reductions considered earlier. In the Wick-rotated $D=4$ theory
obtained {\it via} an $S^7$ reduction, we obtain an instanton that
shares some features with the Hawking-Turok instanton \cite{ht1}, but
now obtained as a specific solution of Euclideanised $D=11$
supergravity. Moreover, we shall see that one can also follow the
spirit of Ref.\ \cite{bous} and combine domain walls and instantons
together in a nonsingular solution with both Euclidean and Minkowskian
regions. Although we have not carried out a full analysis of the
cosmological implications of such solutions, we note that the
occurrence of domain walls and instantons within these dimensionally
reduced supergravities appears to provide some of the elements
postulated in recent cosmological discussions, but now within the
definite context of supergravity theory.

\section{Sphere reductions of supergravities\label{sec:1}}

     In this section, we consider the consistent reductions of various
supergravity theories on spheres, using the general results obtained
in appendix \ref{app:a}.  Specifically, we shall consider the cases of
$D=11$ supergravity compactified on $S^7$ and on $S^4$, type IIB
supergravity compactified on $S^5$, $D=6$ supergravity on $S^3$, and
$D=5$ supergravity on $S^2$.  In all the cases where the internal
space is a sphere of odd dimension, one can also retain a further
singlet mode in a consistent truncation, namely the ``squashing'' mode
parameterising the length of the $U(1)$ fibres in the description of
$S^{2m+1}$ as a $U(1)$ bundle over $CP^m$.  In this case, the modes
that are retained will be singlets under the $SU(m+1)\times U(1)$
subgroup of $SO(2m+2)$ that is the isometry group of the generic
squashed sphere.  As discussed in appendix \ref{app:a}, the results
are also easily generalised to the case of timelike reductions on AdS
spaces instead of spheres.  In these cases the breathing-mode potential in the
lower-dimensional Lagrangian is reversed in sign.  Note also that all
the results for sphere and AdS reductions apply equally well to any
other internal Einstein spaces, with positive, negative or zero Ricci
scalar.

\subsection{Reduction of $D=11$ supergravity on $S^7$\label{ssec:1.1}}

     Following the general discussion in appendix \ref{app:a}, we
reduce the eleven-dimensional fields $\hat g_{MN}$ and $\hat F_4$ on
$S^7$:
\bea
d\hat s^2 &=& e^{2\alpha\varphi}\, ds_4^2 + e^{2\beta\varphi}\, ds_7^2
\ ,\nn\\ 
\hat F_\4 &=& F_\4\ .\label{d11d4red}
\eea
Here, the metric $ds_4^2$ and the fields $F_\4$ and $\varphi$ depend
only on the four coordinates of the four-dimensional space, whereas
the hatted fields depend, {\it a priori}, on all eleven coordinates.
The metric $ds_7^2$ is a fixed metric on the round 7-sphere, with a
fixed standard radius.  The constants $\a$ and $\b$ are given by
\be
\alpha=\ft{\sqrt7}{6}\ ,\qquad \beta = -\ft27 \alpha\ .
\ee
We then obtain the $D=4$ Lagrangian 
\be
e^{-1}{\cal L} = R -\ft12(\del\varphi)^2 + 
e^{\ft{18\a}{7}\varphi}\, R_7 - \ft{1}{48} e^{-6\a\varphi}\,
F_\4^2\ ,\label{s7lag}
\ee
Here, $R_7$ is the Ricci scalar of the 7-sphere; it is just a fixed
constant.  

     The four-dimensional field equations following from this
Lagrangian are
\bsea
&&R_{\mu\nu} = \ft12\del_\mu\varphi\, \del_\nu\varphi + 
\ft1{12} e^{-6\a\varphi}\, (F^2_{\4 \mu\nu} -\ft38 F_\4^2\, g_{\mu\nu})
-\ft12 e^{\ft{18\a}{7}\varphi}\, R_7\, g_{\mu\nu} \ , \\
&&\square\varphi = -\ft{18\a}{7}\, e^{\ft{18\a}{7}\varphi}\, R_7 -
 \ft{\a}{8}\, e^{-6\a\varphi}\, F_\4^2\ ,\label{feq}\\
&&\nabla^\mu(e^{-6\a\varphi}\, F_{\mu\nu\rho\sigma}) = 0\ .
\esea
The last equation implies that $F_4$ can be solved by writing
\be
F_{\mu\nu\rho\sigma} = c\, e^{6\a\varphi} \epsilon_{\mu\nu\rho\sigma}\ ,
\label{f4dual}
\ee
where $c$ is a constant.  Either by substituting this back into the
field equations, or by following the standard procedure for dualising
a field, we can obtain the Lagrangian describing the dualised system
\be
e^{-1}{\cal L} = R -\ft12(\del\varphi)^2 - V(\varphi)\ ,\label{s7lag2}
\ee
where the potential $V(\varphi)$ is given by
\be
V(\varphi) = \ft12 c^2 e^{6\a\varphi} - e^{\ft{18\a}{7}\varphi}\, R_7
\ .\label{phipot}
\ee      
The Lagrangian (\ref{s7lag2}) is a special case of (\ref{emlag}) given
in appendix \ref{app:a}.

     It can be seen that this $D=4$ theory admits a solution where
$\varphi=\varphi_{\sst{\rm AdS}}$ is a constant.  The Ricci scalar
$R_7$ of the 7-sphere is then given in terms of $c$ and
$\varphi_{\sst{\rm AdS}}$.  This includes the AdS$_4 \times S^7$
solution.  Specifically, we will have:
\bea
R_{\mu\nu} &=& -\ft13 c^2\, e^{6\a\varphi_{\sst{\rm AdS}}}\, 
g_{\mu\nu}\ ,\nn\\
e^{\ft{24\a}{7}\varphi_{\sst{\rm AdS}}} &=& \fft{6R_7}{7 c^2}
\ .\label{adsphi}
\eea
Later in the paper, we shall obtain a domain-wall solution of this
theory that interpolates between this ``vacuum'' solution and flat
space.  We shall also show that this four-dimensional theory admits
cosmological instanton solutions, similar to those described in
\cite{ht1}.

\subsection{Inclusion of a squashing mode in the $S^7$ reduction
\label{ssec:1.2}}

     Another consistent truncation of the $S^7$ reduction from $D=11$
can be obtained by retaining all the singlets under the $SU(4)\times
U(1)$ subgroup of $SO(8)$, rather than just the SO(8) singlets.  Since
$SU(4)\times U(1)$ is the isometry group of the one-parameter family
of 7-spheres constructed as $U(1)$ bundles over $CP^3$, for generic
values of the ``squashing parameter'' along the $U(1)$ fibres, it
follows that the $D=4$ scalar field corresponding to this squashing
mode is a singlet under $SU(4)\times U(1)$. In fact, we can treat this
problem by viewing the reduction on $S^7$ as a reduction first to
$D=10$ type IIA, followed by a reduction on $CP^3$, in which we keep
just the singlets under the $SU(4)$ isometry group of $CP^3$.  Thus
the full set of fields in $D=4$ that are singlets under $SU(4)\times
U(1)$ will be the direct reduction of the fields already present in
$D=10$ type IIA, plus the further scalar field corresponding to the
``breathing mode'' of $CP^3$.\footnote{A similar kind of analysis for
the squashing of $S^7$ described as an $SU(2)$ bundle over $S^4$,
which arises in the $N=1$ supersymmetric AdS$_4\times S^7$ ``squashed
seven-sphere'' compactification of $D=11$ supergravity \cite{adp} can
be found in \cite{page}.}

     To establish notation, let us write the ansatz for the $D=11$
metric reduced to $D=10$ as
\be
ds_{11}^2 = e^{-\ft16\phi}\, ds_{10}^2 + e^{\ft43\phi}\, (dz+{\cal A})^2
\ .\label{d11d10metric}
\ee
Note that $\phi$ is nothing but the dilaton of the type IIA theory.
In this convention, the ten-dimensional string coupling constant is
given by $\lambda_{10} = e^{\phi}$.  Then the reduction on 
$CP^3$ from $D=10$ to $D=4$ will be performed as follows:
\be
ds_{10}^2 = e^{2\a\varphi}\, ds_4^2 + e^{2\beta\varphi}\, ds^2(CP^3)\ .
\label{d10d4metric}
\ee
This part of the reduction proceeds identically to the sphere reductions
discussed in the previous section.  (No special features of the sphere 
metric were needed in that discussion; we can equally well apply the
previous results to reductions on any Einstein metric.)  Thus we 
will have 
\be
\a=\ft{\sqrt3}{4}\ ,\qquad \beta= -\ft13 \a\label{abcp3}
\ee
in this case.  The volume of the $CP^3$ measured in the ten-dimensional
string frame is given by
\be
V_{CP^3} = e^{\ft32 \phi + 6\beta\varphi}\ .
\ee
Thus the four-dimensional string coupling constant is given by
\be
\lambda_4^2 = \fft{\lambda_{10}^2}{V_{CP^3}} =
e^{\ft12 \phi + \ft{\sqrt3}{2}\varphi}\ .\label{d4strcoup}
\ee

     The various fields in $D=10$ will be reduced as follows:
\bea
\hat F_\4 &=& F_\4\ ,\qquad \hat F_\3 = F_\3\ ,\nn\\
\hat{\cal F}_\2 &=& {\cal F}_\2 + 2m\, J\ ,\label{2ared}
\eea
where $J$ is the K\"ahler form on $CP^3$.  In other words, $\hat F_\4$
and $\hat F_\3$ are just taken to be independent of the $CP^3$
coordinates, while $\hat {\cal F}_\2$ is taken to have its background
value as in the AdS$_4\times S^7$ solution (see \cite{swos}), plus a
fluctuation.  The final result for the reduced Lagrangian in $D=4$
will therefore be
\bea
e^{-1}\, {\cal L}&=& R -\ft12(\del\phi)^2 -\ft12(\del\varphi)^2 
+  e^{\ft83\a\varphi}\, R_6 -6m^2\, e^{\ft32\phi+\ft{10\a}{3}\varphi}
\nn\\
&&-\ft1{48} e^{\ft12\phi-6\a\varphi}\, F_\4^2
-\ft1{12}e^{-\phi -4\a\varphi}\, F_\3^2 -\ft14 e^{\ft32\phi-2\a\varphi}
\, {\cal F}_\2^2\ ,\label{cp3lag}
\eea
where $R_6$ is the Ricci scalar of the compactifying $CP^3$ space.

     A simple check is to verify that we can recover the usual 
AdS$_4\times S^7$ solution of $D=11$ supergravity.  Now,  
from (\ref{cp3lag}), we have that the equations of motion for the two
dilatons and the metric are
\bea
\square\phi &=&9m^2\, e^{\ft32\phi+\ft{10\a}{3}\varphi} + 
\ft1{96} e^{\ft12\phi-6\a\varphi}\, F_\4^2 -\ft1{12}
e^{-\phi-4\a\varphi} \, F_\3^2 +
\ft38 e^{\ft32\phi-2\a\varphi}\, {\cal F}_\2^2 \ ,\nn\\
\square\varphi &=& -\ft83\a\, R_6 \, e^{\ft{8\a}{3}\varphi}\, 
+20\a\, m^2\, e^{\ft32\phi+\ft{10\a}{3}\varphi}
- \ft{\a}{8} e^{\ft12\phi-6\a\varphi}\, F_\4^2 \nn\\
&&- \ft{\a}{3}\, e^{-\phi-4\a\varphi} \, F_\3^2 -
\ft{\a}{2} e^{\ft32\phi-2\a\varphi}\, {\cal F}_\2^2 
\ ,\nn\\
R_{\mu\nu} &=& \ft12\del_\mu\phi\, \del_\nu\phi +
               \ft12\del_\mu\varphi\, \del_\nu\varphi-
               \ft12 R_6\,  e^{\ft{8\a}{3}\varphi}\, g_{\mu\nu} +
           3m^2\,  e^{\ft32\phi+\ft{10\a}{3}\varphi}\, g_{\mu\nu} \nn\\
        &&+\ft1{12} e^{\ft12\phi-6\a\varphi}\, (F_{\4 \mu\nu}^2 -
                        \ft38 F_\4^2\, g_{\mu\nu}) +
         \ft14 e^{-\phi-4\a\varphi}\, (F_{\3\mu\nu}^2 -\ft13 F_\3^2\,
                                  g_{\mu\nu} )\nn\\
&& +\ft12e^{\ft32\phi-2\a\varphi}\, ({\cal F}_{\2\mu\nu}^2 -
                                \ft14{\cal F}_\2^2\, g_{\mu\nu})\ .
\eea
We see that there is indeed a solution where $F_\4 = c\, e^{-\ft12\phi
+ 6\a\varphi}\, \epsilon_\4$, and with $\phi=\phi_0$ and
$\varphi=\varphi_0$ constants given by
\be
e^{2\phi_0} = \sqrt{\fft{c}{6m}}\, \fft{R_6}{48m^2}\ ,\qquad
e^{\ft{16\a}{3}\varphi_0} = \fft{3 R_6^2}{32m\, c^2}\ .
\ee
The Ricci tensor satisfies
\be
R_{\mu\nu} = -\ft14 R_6\, e^{\ft{8\a}{3}\varphi_0}\, g_{\mu\nu}\ ,
\ee
which admits, in particular, the standard AdS$_4$ solution. 

    Another check is to see that we can obtain the $SO(8)$-invariant
truncation of section \ref{ssec:1.1} in a consistent way.  To do this,
we can perform a rotation of the dilatons in the case above, and
define a pair $(\tilde\phi,\tilde\varphi)$ as follows:
\bea
\tilde\phi &=& \ft{3\sqrt3}{2\sqrt7}\, \phi +
               \ft1{2\sqrt7}\, \varphi\, \nn\\
\tilde\varphi &=& -\ft1{2\sqrt7}\, \phi + \ft{3\sqrt3}{2\sqrt7}\, 
                        \varphi\ .\label{redef}
\eea
Substituting into (\ref{cp3lag}), and dropping the fields $F_\3$ and
${\cal F}_\2$, we obtain
\bea
e^{-1}\, {\cal L}&=& R -\ft12(\del\td\phi)^2 -\ft12(\del\td\varphi)^2 
+ R_6\, e^{\ft{3}{\sqrt7}\td\varphi+\ft1{\sqrt{21}}\td\phi} -
     6m^2\, e^{\ft{3}{\sqrt7}\td\varphi+ \ft8{\sqrt{21}}\td\phi }
\nn\\
&&-\ft1{48} e^{-\sqrt7\td\varphi}\, F_\4^2 \ .\label{cp3lag2}
\eea
We see that the equation of motion for $\td\phi$ allows us to set
$\td\phi=0$, provided that $R_6=48m^2$, whereupon the Lagrangian
(\ref{cp3lag2}) reduces to the Lagrangian (\ref{s7lag}) of section
\ref{ssec:1.1} (with $R_7=42m^2$), with $\varphi$ replaced by
$\td\varphi$.  Furthermore, if we write the ansatz
(\ref{d11d10metric},\ref{d10d4metric}) for the eleven-dimensional
metric in terms of the rotated tilde basis for the dilatons, we have
\be
ds_{11}^2 = e^{\ft{\sqrt7}{3}\td\varphi}\, ds_4^2 +
 e^{-\ft2{3\sqrt7}\td\varphi}\, \Big( e^{-\ft1{\sqrt{21}}\td\phi} \,
ds^2(CP^3) + e^{\ft{6}{\sqrt{21}}\td\phi}\, (dz+ {\cal A}_\1)^2\Big)\ .
\ee
Thus we see that indeed the $\td\phi$ scalar describes a
volume-preserving $6+1$ squashing of the $S^7$ metric, while
$\td\varphi$ is the same as the dilatonic scalar $\varphi$ of the
$SO(8)$-invariant ansatz of section \ref{ssec:1.1}.  Thus we see that
the four-dimensional string coupling constant $\lambda_4$, given in
(\ref{d4strcoup}) is not an $SO(8)$-invariant quantity. To get a
four-dimensional theory with AdS$_4$ background, we need to set
$\td\phi=0$, implying that $\lambda_4 = \lambda_{10}^{-2}$
\cite{swos}.

\subsection{$S^4$ reduction of $D=11$\label{ssec:1.3}}

     In this case, the $S^4$ solution requires that the internal
components of $F_4$ (in the 4-sphere) should be non-zero, and so we
must make the following ansatz:
\bea
d\hat s_{11}^2 &=& e^{2\a\varphi}\, ds_7^2 + e^{2\beta\varphi}
\, ds_4^2\ ,\nn\\
\hat F_\4 &=& F_\4 + 6m\, \epsilon_\4 \ .\label{d11d7red}
\eea
 From appendix \ref{app:a}, we see that the constants $\a$ and $\beta$
are given by $\a=\ft2{3\sqrt{10}}$ and $\beta=-\ft54 \a$.  Plugging
into the Lagrangian, we get
\be
{\cal L}_7 = eR -\ft12 e(\del\varphi)^2 + e\, e^{\ft{9\a}{2}\varphi}
\, R_4
 -18 e\, m^2\, e^{12\a\varphi} -\ft1{48} e\, e^{-6\a\varphi}\, F_\4^2 
-3m\, {*(F_\4\wedge A_\3)}\ .\label{d7lag}
\ee
Note that we have a topological mass term for $A_\3$ here, coming from
the $FFA$ term in $D=11$.

     The equations of motion following from this Lagrangian are
\bea
R_{\mu\nu} &=& \ft12 \del_\mu\varphi\, \del_\nu\varphi -\ft15 
e^{\ft{9\a}{2}\varphi}\, R_4\, g_{\mu\nu} + 
\ft{18}{5} m^2\, e^{12\a\varphi} \, g_{\mu\nu}\nn\\
&& \qquad 
+\ft1{12}e^{-6\a\varphi}\, (F_{\mu\nu}^2 -
\ft1{10} F_\4^2\, g_{\mu\nu}) \ ,\nn\\
\square\varphi &=& -\ft92\a\, e^{\ft{9\a}{2}\varphi}\, R_4\, 
                  +216\a\, m^2\, e^{12\a\varphi} -\ft18\a\, 
           e^{-6\a\varphi}\,
              F_\4^2\ ,\label{d7eom}\\
d(e^{-6\a\varphi}\, {*F_\4}) &=& -6m\, F_\4\ .\nn
\eea  
These equations admit an AdS$_7$ solution which corresponds to the
AdS$_7\times S^4$ ``vacuum'' solution of $D=11$ supergravity.
We see from (\ref{d7eom}) that with $F_\4=0$, we can take 
$\varphi=\varphi_{\sst{\rm AdS}}$ to be constant, given by
\be
e^{\ft{15\a}{2}\varphi_{\sst{\rm AdS}}} = \fft{R_4}{48m^2}\ .
\ee
The Ricci tensor in spacetime is then given by
\be
R_{\mu\nu} = -6m^2\, e^{12\a\varphi_{\sst{\rm AdS}}}\, g_{\mu\nu}\ ,
\ee
which allows an AdS$_7$ solution.

\subsection{$S^5$ reduction of type IIB supergravity\label{ssec:1.4}}

     In the case of type IIB supergravity \cite{2b}, we must work at
the level of the $D=10$ equations of motion, since there there is no
Lagrangian formulation of the theory.  The type IIB bosonic equations
of motion can be written as the following \cite{trombone}
\bea
R_{\mu\nu} &=& \ft12\del_\mu\phi\, \del_\nu\phi + \ft12 e^{2\phi}\, 
\del_\mu\chi\, \del_\nu\chi  + \ft1{96} (H_\5)^2_{\mu\nu}  \nn\\
&& +\ft14e^{\phi}\, ((F_\3^1)^2_{\mu\nu} 
-\ft1{12} (F_\3^1)^2\, g_{\mu\nu}) +\ft14e^{-\phi}\, 
((F_\3^2)^2_{\mu\nu} -\ft1{12} (F_\3^2)^2\, g_{\mu\nu})\ ,\nn\\
d({\cal M} {*H_\3}) &=& H_\5 \wedge \Xi\, H_\3\ ,\label{2beom}\\
H_\5 &=& {* H_\5}\ ,\nn\\
d(e^{2\phi} {*d\chi}) &=& - e^{\phi}\, F_\3^2 \wedge F_\3^1\ ,\nn\\
d{*d\phi} &=& e^{2\phi} \, d\chi\wedge {*d\chi} + \ft12 e^{\phi}\, 
F_\3^1 \wedge F_\3^1 - \ft12 e^{-\phi}\, F_\3^2\wedge F_\3^2\ ,
\eea
where 
\be
{\cal M} = \pmatrix{e^\phi & \chi\, e^{\phi}\cr
               \chi\, e^{\phi} & e^{-\phi} +\chi^2\, e^\phi}\ ,\qquad
H_\3 = \pmatrix{dA_\2^1\cr dA_\2^2}\ ,\qquad 
\Xi=\pmatrix{0&1\cr -1&0}\ .
\ee
The self-dual 5-form $H_\5$ satisfies the Bianchi identity $dH_\5 +
\ft12 \epsilon_{ij}\, F_\3^i\wedge F_\3^j=0$.  The R-R field strength
$F_\3^1$ is given by $F_\3^1 = dA_\2^1 -\chi\, dA_\2^1$, and the NS-NS
field strength is given by $F_\3^2 = dA_\2^2$.

     The Kaluza-Klein ansatz for the metric will be the usual one,
invariant under the isometry group $SO(6)$ of the compactifying
5-sphere:
\be
ds_{10}^2 = e^{2\a\varphi}\, ds_5^2 + e^{2\beta\varphi}\, ds^2(S^5)\ .
\ee
 From the general results in appendix \ref{app:a}, we have here that
\be
\a = \ft{1}{4}\sqrt{\ft53}\, \qquad \beta = -\ft35\, \a\ .
\ee
The ansatz for the self-dual 5-form, which is non-dynamical 
in the reduced $D=5$ theory, will be
\be
H_\5 = 4m\, e^{8\a\varphi}\, \epsilon_\5 + 4m\, \epsilon_\5(S^5)\ ,
\ee
where the $\epsilon_\5$ quantities are the volume forms on the
five-dimensional spacetime and the $S^5$ metrics $ds_5^2$ and
$ds_5^2(S^5)$ respectively.  The choice of ansatz here is dictated by
the requirements that the 5-form be self-dual, and that it satisfy the
necessary Bianchi identity $dH_\5=0$.  After some algebra, we find
that substituting these ans\"atze into the ten-dimensional equations of
motion leads consistently to the the 5-dimensional equations
\bea
R_{\mu\nu} &=& \ft12\del_\mu\phi\, \del_\nu\phi + \ft12 e^{2\phi}\,
\del_\mu\chi\, \del_\nu\chi + \ft83\, m^2 e^{8\a\varphi} \, g_{\mu\nu}
-\ft{1}{3} e^{\ft{16\a}{5}\varphi}\, R_5 \, g_{\mu\nu} \nn\\
&& +\ft14e^{\phi}\, ((F_\3^1)^2_{\mu\nu} -\ft2{9} (F_\3^1)^2\, g_{\mu\nu})
+\ft14e^{-\phi}\,
((F_\3^2)^2_{\mu\nu} -\ft2{9} (F_\3^2)^2\, g_{\mu\nu})\ ,\nn\\ 
\square\varphi &=& 64\a\, m^2\, e^{8\a\varphi} -
\ft{16}{5} \a\, e^{\ft{16\a}{5}\varphi}\, R_5-
\ft13\a\, e^{-\phi-4\a\varphi}\, (F_\3^1)^2
-\ft13\a\, e^{\phi-4\a\varphi}\, (F_\3^2)^2\ .\hspace{.5cm}
\eea
(The other equations of motion do not immediately concern us here.)  It is
not hard to see that the full set of equations of motion can be derived
from the 5-dimensional Lagrangian
\bea
{\cal L}_5 &=& eR -\ft12 e\, (\del\phi)^2 -\ft12 e\, (\del\varphi)^2
-\ft12 (\del\chi)^2 e^{2\phi} 
-8m^2\, e\, e^{8\a\varphi} +  e^{\ft{16\a}{5}\varphi}\, R_5 \nn\\
&&-\ft1{12} e\, e^{\phi-4\a\varphi}\, (F_\3^1)^2 
-\ft1{12} e\, e^{-\phi-4\a\varphi}\, (F_\3^2)^2 -2 m\, {*(\epsilon_{ij}\,
A_\2^i\wedge dA_\2^j)}\ .\label{d52blag1}
\eea
Note that this reduction allows the expected AdS$_5$ solution, with
$\varphi=\varphi_{\sst{\rm AdS}}$ being constant and given by
\be
e^{\ft{24\a}{5}\varphi_{\sst{\rm AdS}}} = \fft{R_5}{20 m^2}\ .
\ee
The Ricci tensor for AdS$_5$ is then given by
\be
R_{\mu\nu} = -4m^2\, e^{8\a\varphi_{\sst{\rm AdS}}}\, g_{\mu\nu}\ .
\ee

\subsection{Inclusion of a squashing mode in the $S^5$ 
  reduction\label{ssec:1.5}}

    In the same manner as we did previously for $S^7$, we can include
a squashing mode in the reduction of Type IIB supergravity on $S^5$,
in which we view $S^5$ as a $U(1)$ bundle over $CP^2$, and thus we now
truncate so as to keep all the $SU(3)\times U(1)$ singlets in the
decomposition of the $SO(6)$ isometry group of the round $S^5$ under
the $SU(3)\times U(1)$ subgroup that is the symmetry group of the
squashed family of $S^5$ metrics.  We may use a similar trick as in
section \ref{ssec:1.2}, and carry out the process by first considering
the $S^1$-reduced $D=9$ metric, and then reducing this on $CP^2$,
while giving the appropriate background value to the Kaluza-Klein
2-form field strength in $D=9$.  We shall express the fields of the
$D=9$ theory in the type IIB language, since our aim is to describe
the squashed-$S^5$ reduction of the Type IIB theory.

     The $D=9$ Lagrangian, obtained from the type IIB theory by
reducing the metric according to the usual ansatz
\be
ds_{10}^2 = e^{-\ft1{2\sqrt7}\varphi}\, ds_9^2 +
   e^{\ft{\sqrt7}{2}\varphi}\, (dz+ {\cal A}_\1)^2 \label{d10d9red}
\ee
 is
\bea
e^{-1} {\cal L}_{9} &=& R-\ft12 (\del\phi)^2 -\ft12 (\del
\varphi)^2 - \ft12 e^{2\phi} (\del \chi)^2 \nonumber\\
&&-\ft1{48} e^{-\ft2{\sqrt7}
\varphi} F_\4^2 -\ft1{12} e^{-\phi+\ft1{\sqrt7}\varphi} (F_\3^{({\rm
NS})})^2 -\ft12 e^{\phi +\ft1{\sqrt7} \varphi} (F_\3^{({\rm R})})^2
\label{d92blag}\\
&&-\ft14 e^{\ft4{\sqrt7} \varphi} ({\cal F}_\2)^2 -
\ft14 e^{\phi - \ft3{\sqrt7}\varphi} (F_\2^{({\rm R})})^2 -
\ft14 e^{-\phi - \ft3{\sqrt7} \varphi} (F_\2^{({\rm NS})})^2\nonumber\\
&&-\ft1{2e} \, dA_\3 \wedge dA_\3 \wedge {\cal A}_\1 +
\ft1{e}\, dA_\2^{({\rm NS})} \wedge dA_\2^{({\rm R})} \wedge A_\3\ .
\nn
\eea
The reduction ansatz for the metric will be the usual one, namely
\be
ds_9^2 = e^{2\a f} \, ds_5^2 + e^{2\beta f}\, ds_4^2(CP^2)
\ ,\label{d9d5red}
\ee
where, from appendix \ref{app:a}, the constants $\a$ and $\beta$ here
are 
given by
\bea
\a = \sqrt{\ft2{21}}\ ,\qquad \beta = -\ft34\, \a\ .
\eea
For the various field strengths, they will all just directly reduce
na\"{\i}vely, except for $F_\4$ and ${\cal F}_\2$, which will be reduced
according to
\bea
F_\4 &\longrightarrow & F_\4 +4m\, \epsilon_\4\ ,\nn\\
{\cal F}_\2 & \longrightarrow & {\cal F}_\2 + 2\mu\, J\ ,\label{cp2fred}
\eea
where $\epsilon_\4$ is the volume-form on $CP^2$, and $J$ is the
K\"ahler form.  

    We may now substitute the various ans\"atze into the
nine-dimensional Lagrangian (\ref{d92blag}).  To avoid repetitive
formulae, let us also at this stage perform an $O(2)$ rotation of the
scalars $(\varphi, f)$, to tilded ones:
\bea
&&\td f = \sqrt{\ft{3}{35}}\, f + 4 \sqrt{\ft2{35}}\, \varphi\ ,\qquad
\td\varphi =  4 \sqrt{\ft2{35}}\, f - \sqrt{\ft{3}{35}}\, \varphi\ , \nn\\
&&f = \sqrt{\ft{3}{35}}\, \td f + 4 \sqrt{\ft2{35}}\, \td\varphi\ ,\qquad
\varphi =  4 \sqrt{\ft2{35}}\, \td f - \sqrt{\ft{3}{35}}\,
\td\varphi\ . \label{d5redef}
\eea
In terms of these, the reduced five-dimensional Lagrangian turns out
to be
\bea
e^{-1} {\cal L}_{5} &=& R-\ft12 (\del\phi)^2 -\ft12 (\del
\td\varphi)^2 -\ft12 (\del\td f)^2 - \ft12 e^{2\phi} (\del \chi)^2 
-\ft1{48} e^{-2\sqrt{\ft25}\td f-2\sqrt{\ft35}\td\varphi} F_\4^2 \nn\\
&&-\ft1{12} e^{-\phi -\sqrt{\ft53}\td\varphi}\, (F_\3^{({\rm
NS})})^2 -\ft12 e^{\phi -\sqrt{\ft53}\td \varphi}\, (F_\3^{({\rm R})})^2
-\ft14 e^{2\sqrt{\ft25}\td f -\ft4{\sqrt{15}}\td\varphi}\, 
({\cal F}_\2)^2\nn\\
&& - \ft14 e^{\phi - 2\sqrt{\ft25} \td f -\ft1{\sqrt{15}}\td\varphi}\, 
(F_\2^{({\rm R})})^2 -
\ft14 e^{-\phi - 2\sqrt{\ft25} \td f -\ft1{\sqrt{15}}\td\varphi}\, 
(F_\2^{({\rm NS})})^2\label{d52blag2}\\
&& -8 m^2\, e^{2\sqrt{\ft53}\td\varphi} -
\mu^2\, e^{3\sqrt{\ft25}\td f +\ft4{\sqrt{15}} \td\varphi} +
e^{\ft1{\sqrt{10}}\td f + \ft4{\sqrt{15}} \td\varphi}\, R_4 \nn \\
&&- \ft4{e}\, m \, {*( dA_\3 \wedge {\cal A}_\1)} -\ft4{e}\, m\,
{*(A_\2^{({\rm R})} \wedge dA_\2^{({\rm NS})})}\ .
\nn
\eea
     
     We can first perform the statutory consistency check of verifying
that we can get back the results of the previous subsection, by
truncating out the squashing mode.  The rotated basis for the dilatons
that we are using in (\ref{d52blag2}) is adapted for the purpose, and
indeed we can see it is consistent with the equation of motion for
$\td f$ to set $\td f=0$, provided that $R_4=6\, \mu^2$.  If we then
set $\td f=0$, we see that indeed the Lagrangian truncates to the
previous one (\ref{d52blag1}), with $\varphi$ replaced by $\td
\varphi$.

    It is also instructive to write the ansatz for the original
ten-dimensional type IIB metric in terms of the tilded dilatons
defined in (\ref{d5redef}).  From (\ref{d10d9red}) and
(\ref{d9d5red}), we find that the ten-dimensional metric is given by
\be
ds_{10}^2 = e^{\sqrt{\ft{5}{12}}\td\varphi}\, ds_5^2 +
e^{-\sqrt{\ft{5}{12}}\td\varphi}\, \Big( e^{-\ft1{\sqrt{10}}\td f}\, 
   ds_4^2(CP^2) + e^{\ft4{\sqrt{10}}\td f}\, (dz+{\cal A}_\1)^2\Big) \ .
\ee
Thus indeed we see that $\td f$ acts as a volume-preserving
``squashing mode'' of the 5-sphere, while $\td\varphi$ is the
breathing mode.

\subsection{Further examples\label{ssec:1.6}}

\subsubsection{Reduction of $D=6$ supergravity on $S^3$\label{sss:d6red}}

    We may consider the reduction of a $D=6$ Lagrangian of the form 
\be
e^{-1}\, {\cal L}_6 = R -\ft12(\del\phi)^2 -\ft1{12} e^{\sqrt2\phi}\,
F_\3^2\ ,\label{d6lag}
\ee
where the internal space is taken to be $S^3$.  This calculation is
relevant for the reduction of any of the six-dimensional supergravities.
Following the same procedures that we have used previously, we shall
first consider the situation where an $SO(4)$-invariant reduction is
performed, with the ans\"atze
\bea
d\hat s_6^2 &=& e^{2\a\varphi}\, ds_3^2 + e^{2\beta\varphi}\, ds^2(S^3)\ ,
\nn\\
\hat F_\3 &=& F_\3 + m\, \epsilon_\3(S^3)\ ,\label{d6red1}
\eea
where $\a=\sqrt{\ft{3}{8}}$ and $\beta=-\ft13\a$.  The resulting 
three-dimensional Lagrangian is
\be
e^{-1}\, {\cal L}_3 = R -\ft12(\del\phi)^2 -\ft12(\del\varphi)^2 
+ R_3\, e^{\ft{8\a}{3}\varphi} -\ft12 m^2\, e^{\sqrt2\phi+4\a\varphi}
 -\ft1{12} e^{\sqrt2\phi-4\a\varphi}\, F_\3^2\ .
\ee
We may now dualise $F_\3$ in the standard way, solving its equation of
motion by writing $F_\3=c\, e^{-\sqrt2\phi+4\a\varphi}\, 
\epsilon_\3(S^3)$.  The resulting dualised Lagrangian is
\be
e^{-1}\, {\cal L}_3 = R -\ft12 (\del\phi)^2 -\ft12(\del\varphi)^2
+ R_3\, e^{\ft{8\a}{3}\varphi} -\ft12 m^2 \, e^{\sqrt2\phi+4\a\varphi}
-\ft12 c^2\, e^{-\sqrt2\phi+4\a\varphi}\ .\label{d3lag}
\ee

     As in the previous examples, we may also consider a more general
reduction where the squashing mode of the 3-sphere is included,
corresponding to the family of $SO(3)\times U(1)$ invariant metrics on
the 3-sphere described as a $U(1)$ bundle over $CP^1=S^2$.  As before,
this is most easily done by first reducing the six-dimensional theory
on a circle, which will then be taken to be the $U(1)$ of the Hopf
fibration.  Thus our starting point now will be the five-dimensional
theory obtained from (\ref{d6lag}) by reducing the metric according to
\be
d\hat s_6^2 = e^{-\ft1{\sqrt6}\varphi}\, ds_5^2 +
e^{\ft3{\sqrt6}\varphi}\, (dz+{\cal A}_\1)^2\ ,
\ee
which gives the Lagrangian
\bea
e^{-1}\, {\cal L}_5 &=& R -\ft12(\del\phi)^2 -\ft12(\del\varphi)^2
-\ft1{12} e^{\sqrt2\phi+\ft2{\sqrt6}\varphi}\, F_\3^2 \nn\\
&&-\ft14 e^{\sqrt2\phi-\ft2{\sqrt6}\varphi}\, F_\2^2 
-\ft14 e^{\ft4{\sqrt6}\varphi}\, {\cal F}_\2^2\ .
\eea
We then perform a further reduction on $S^2$, with the ans\"atze
\bea
ds_5^2 &\rightarrow& e^{\ft2{\sqrt3} f}\, ds_3^2 + e^{-\ft1{\sqrt3}f}\, 
ds^2(S^2)\ ,\nn\\
F_\3 &\rightarrow& F_3\ ,\nn\\
F_\2 &\rightarrow& F_\2 + \lambda\, J\ ,\nn\\
{\cal F}_\2 &\rightarrow& {\cal F}_\2 + m\, J\ .
\eea
It is convenient also to make an orthogonal transformation on the 
basis for the dilatons $\varphi$ and $f$, defining
\be
f=\ft{2\sqrt2}{3} \td\varphi + \ft13 \td f\ ,\qquad 
\varphi= -\ft13\td\varphi + \ft{2\sqrt2}{3}\td f\ .
\ee
In this basis, and after dualising $F_\3$ in $D=3$ by solving its
equation of motion in the form $F_\3 = c\, e^{-\sqrt2\phi -
\ft2{\sqrt6}\varphi +4\a f}$, we obtain the three-dimensional 
Lagrangian
\bea
e^{-1}\, {\cal L}_3 &=& R -\ft12(\del\phi)^2 -\ft12(\del\td\varphi)^2
-\ft12 (\del\td f)^2 -\ft14 e^{\sqrt2\phi -\ft2{\sqrt3}\td f -
\sqrt{\ft23}\td\varphi}\, F_\2^2 -\ft14 e^{\ft2{\sqrt3}\td f -
2\sqrt{\ft23}\td\varphi}\, {\cal F}_\2^2\hspace{.5cm}\nn\\
&&-\ft12\lambda^2\, e^{\sqrt2\phi +\sqrt6\td\varphi} -\ft12 c^2\, 
e^{-\sqrt2\phi+\sqrt6\td\varphi} -\ft12 m^2\, e^{\ft4{\sqrt3}\td f+
2\sqrt{\ft23}\td\varphi} + R_2\, e^{\ft1{\sqrt3}\td f + 2\sqrt{\ft23}
\td\varphi}\ .
\eea
As usual, we can verify that the equation of motion for $\td f$ can be
satisfied by setting $\td f=0$, provided that $R_2=2m^2$ and that
$F_\2={\cal F}_\2=0$.  This corresponds to truncating the theory to
the previous $SO(4)$-invariant reduction on $S^3$.  The fact that $\td
f$ is the volume-preserving squashing mode can be seen from the metric
reduction ansatz, expressed in the tilded variables:
\be
d\hat s_6^2 = e^{\sqrt{\ft32}\td\varphi}\, ds_3^2 +
e^{-\ft1{\sqrt6}\td\varphi}\, \Big( e^{-\ft1{\sqrt3}\td f}\, 
ds^2(S^2) + e^{\ft2{\sqrt3}\td f}\, (dz+{\cal A}_\1)^2 \Big)\ .
\ee

\subsubsection{Reduction of $D=5$ supergravity on $S^2$\label{sssec:d5red}.}

    To discuss this, it suffices to consider the subsector of the
$D=5$ theory described by the Einstein-Maxwell Lagrangian
\be
e^{-1}\, {\cal L}_5 = R - \ft14 \hat F_\2^2\ .\label{einmax}
\ee
We then reduce on $S^2$, with the ansatz
\be
d\hat s_5^2 = e^{2\a\varphi}\, ds_3^2 + e^{2\beta\varphi}\, 
ds^2(S^2)\ ,\qquad
\hat F_\2 = F_\2 + m\, \epsilon_\2(S^2)\ ,
\ee
where $\a=1/\sqrt3$ and $\beta=-\a/2$.  The resulting three-dimensional
Lagrangian turns out to be
\be
e^{-1}\, {\cal L}_3 = R -\ft12(\del\varphi)^2 -\ft14 e^{-2\a\varphi}\,
F_\2^2 -\ft12 m^2\, e^{4\a\varphi} + R_2\, e^{3\a\varphi}\ .\label{d3lag2}
\ee

\section{Supersymmetric domain walls\label{sec:2}}

\subsection{Domain wall in $D=4$\label{ssec:2.1}}

        The Lagrangian (\ref{s7lag}) of the $SO(8)$-singlets in the
$S^7$ reduction admits a supersymmetric domain wall (\ie membrane)
solution in $D=4$.  For convenience, we shall discuss the solutions
using the Lagrangian (\ref{s7lag2}), where the $F_4$ term has been
dualised into a cosmological term.  Comparing this with the general
Lagrangian (\ref{domainlag1}, \ref{domainlag2}) in appendix
\ref{app:b}, we see that $a_1=6\a$, $a_2=\ft{18}{7}\a$, $g_1=c$,
$g_2=\sqrt{2R_7}$, and $\lambda=0$, which is consistent with the
requirement (\ref{lambdasquare}).  The solution is therefore given by
$\td b_1= \pm 5c/(2k)$, $\td b_2 =\pm 5\sqrt{6R_7/7}\, /(2k)$,
$\nu_1=\ft12(a_2-a_1) = \ft5{\sqrt7}$ and $\nu_2=6$, and hence
\bea
e^{-\ft5{\sqrt7}\varphi} &=& H = e^{-\ft5{\sqrt7}\varphi_0} 
+ k\, |y|\ ,\nn\\
e^{3A} &=&e^{-B}= \td b_1\, H^{\fft3{10}} + \td  b_2\, H^{\fft7{10}} 
\ ,\label{d4domainsol}
\eea
with the metric of the form given by (\ref{gendomainmet}):
\be
ds_4^2 = \Big(\td b_1\, H^{\fft3{10}} + \td  b_2\,
H^{\fft7{10}}\Big)^{\ft23}\, dx^\mu\, dx_\mu + 
\Big(\td b_1\, H^{\fft3{10}} + \td  b_2\, H^{\fft7{10}}\Big)^{-2}\,
dy^2\ .\label{domwall}
\ee
Note that here we have four different solutions, depending on the
signs of $\td b_i$.  We shall show presently that these solutions
preserve one half of the supersymmetry.  We shall consider the case
where these two parameters are of opposite sign, in order to have
solutions that are real.  In this case, when $k\rightarrow 0$, the
solution reduces to AdS$_4$, with $e^{\ft{24\a}{7}\varphi}
=e^{\ft{24\a}{7} \varphi_{\sst{\rm AdS}}}$ (see appendix \ref{app:b}
for a general derivation of domain-wall solutions).

           We shall first consider the case where the constant
$\varphi_0$ in the solution (\ref{d4domainsol}) is given by
$\varphi_0=\varphi_{\sst{\rm AdS}}$, implying that the geometry near
$y=0$ approaches AdS$_4$.  To see this, we note
that in the region $|y|\rightarrow 0$, the metric can be written as
$ds^2 \sim e^{2\rho/3}\, dx^\mu dx_\mu + d\rho^2$, with $\rho = \log
|y|\rightarrow -\infty$.  This is AdS$_4$, written in horospherical
coordinates.  Note that if $k$ is positive, then the solution is real
for all values of $y$, provided that $\td b_2>0$ and $\td b_1 <0$. In
this case we can take $y$ to run from $-\infty$ to $\infty$.  In the
regions where $|y|\rightarrow \infty$ the $H^{\fft7{10}}$ term in
(\ref{d4domainsol}) dominates.  The curvature tends to zero in these
regions, so the metric is asymptotically locally flat, and its
behaviour is dominated by the contribution from the $R_7$ potential
term (\ie the $H^{\fft 7{10}}$ term), while the dilaton approaches
negative infinity in this asymptotic limit.  The solution is
reflection-symmetric about $y=0$, where the solution approaches
AdS$_4$.  It deviates more and more from AdS$_4$ as $|y|$ increases,
with the curvature eventually vanishing at $|y| =\infty$.  The $y=0$
point is a horizon, since $g_{00}=0$ there.  If $k$ is taken to be
negative instead, then the solution is real for $|y|<y_0\equiv
e^{-\ft5{\sqrt7} \varphi_{\sst{\rm AdS}}}/|k|$.  In the regions
$|y|\rightarrow y_0$, where $H$ vanishes, the curvature becomes
singular and its behaviour is then dominated by the $c^2$ potential
term (\ie the metric is dominated by the term $H^{\fft3{10}}$ in
(\ref{d4domainsol}).) The dilaton approaches positive infinity in this
limit. Thus, in this $k<0$ case, the solution is again
reflection-symmetric, and approaches AdS$_4$ at $y=0$, but it now has
curvature singularities at $y=\pm y_0$.  In both the positive and
negative $k$ cases, the metric functions $A$ and $B$ can be expressed
as \be e^{3A}=e^{-B}=\fft{5}{2k}\Big( \sqrt{6R_7/7}\, H^{\fft7{10}} -
c\, H^{\fft3{10}} \Big)\ , \ee corresponding to $b_1 <0$ and $b_2>0$,
where $b_i$ is defined in (\ref{firstorder}).

     Different situations can arise if the constant $\varphi_0$ takes
values other than $\varphi_{\sst{\rm AdS}}$.  In the case $k>0$, for
which $y$ runs from $-\infty$ to $\infty$, then if
$\varphi_0>\varphi_{\sst\rm AdS}$ the metric reaches the AdS form at
the points $y_\pm = \pm k^{-1}\, (e^{-5\varphi_{\sst\rm AdS}/\sqrt7 }
- e^{-5\varphi_0/\sqrt7})$.  There is a domain wall at $y=0$, with a
delta-function curvature singularity.  Thus the domain wall divides
the spacetime into two mirror-symmetric regions which each has an
AdS$_4$ as its near-horizon structure.  The domain wall is inside the
horizon.  If, on the other hand $\varphi_0 <\varphi_{\sst\rm AdS}$
then $g_{00}$ never reaches zero, and the domain wall at $y=0$ divides
the spacetime into two regions that never reach the AdS$_4$ form.  The
situation is different if $k<0$, in that now, instead of $y$ running
inwards from a flat region at $|y|=\infty$, it runs from curvature
singularities at $y=\pm y_0$.  Again, depending on whether $\varphi_0$
is greater than or less than $\varphi_{\sst\rm AdS}$, the metric
either passes through the AdS$_4$ region before reaching the domain
wall, or else it reaches the domain wall without encountering an
AdS$_4$ region.  Note that the delta-function curvature singularity at
$y=0$, which arises because the harmonic function is taken to depend
on $y$ through its modulus $|y|$, is absent in the special case
$\varphi_0 = \varphi_{\sst\rm AdS}$ described in the previous
paragraph, but is otherwise generically present.  (For a review of
domain walls in $D=4$ supergravities, see \cite{cs}.)

          Although it might seem a bizarre choice from the $D=4$
perspective, another patching-together of segments of the solution
(\ref{d4domainsol}) is possible. One can match the $k<0$ solution for
$y<0$ onto the $k>0$ solution for $y>0$. This effectively removes the
absolute value prescription for $y$ in $H$, thus removing also the
delta-function curvature singularity at $y=0$. Although the metric
(\ref{d4domainsol}) apparently becomes singular at $y=0$, this proves
to be just a coordinate singularity, and the spacetime has in fact a
regular horizon there. The price to be paid for this choice of
patching is the appearance of a genuine curvature singularity at
$y=-y_0$, which can be reached by a lightlike or timelike geodesic at
a finite affine-parameter value. From a $D=4$ viewpoint, the natural
choice would seem to be the first one made above: taking $H$ to depend
on $|y|$ with $k>0$, and tolerating a rather mild delta-function
singularity at $y=0$. We shall shortly see, however, that the
``patched'' solution in fact corresponds directly to the most natural
metric from the $D=11$ point of view. We shall return in the next
subsection to a more general study of the spherical oxidations of
domain-wall solutions.
            
          The general solution can be straightforwardly oxidised back
to $D=11$, giving 
\bea 
ds_{11}^2 &=& e^{2\a\varphi + 2A}\, dx^\mu
dx_\mu + e^{2\a\varphi -6A}\, dy^2 + e^{2\beta\varphi}\, ds_7^2 \nn\\
&=&(\td b_1 H^{-\ft25} + \td b_2)^{\ft23}\, dx^\mu dx^\nu\eta_{\mu\nu}
+ (\td b_1 H^{\fft8{15}} + \td b_2 H^{\fft{14}{15}} )^{-2}\, dy^2 +
H^{\fft2{15}}\, ds_7^2\ ,\nn\\ F_4 &=& c\, e^{6\a\varphi} \epsilon_4=
c\, H^{-\ft75}\, \epsilon_4 = c\, H^{-\ft75}\, d^3x\wedge dy \
.\label{s7d11sol} 
\eea 
 From the eleven-dimensional point of view, the solution has a total
of four free parameters, namely $\varphi_0$, $k$, $c$ and $R_7$.
There are three special limits, depending on the values of these
parameters.  As we have seen, when $k=0$ and $e^{\ft{24\a}{7}\varphi}
=6R_7/(7c^2)$, the solution is AdS$_4\times S^7$.  If $c=0$, the
metric is in fact purely Minkowskian. If $R_7=0$, {\it e.g.}\ when
$ds_7^2$ is a flat metric, then the solution becomes a membrane with
its charges uniformly distributed over the seven-dimensional surface
\cite{lpdomain}.

     It is now fairly simple to check the supersymmetry of the
domain-wall solution.  This is most easily done by looking at the
solution in eleven dimensions, as given in (\ref{s7d11sol}).  First,
we note that a metric of the form
\be
d\hat s^2 = e^{2f}\, dx^\mu\, dx_\mu + e^{2h}\, dy^2 + e^{2u}\, ds^2
\ ,\label{oxdw}
\ee 
where $f$, $h$ and $u$ depend only on $y$, has a spin connection, in
the natural vielbein basis $\hat e^\mu = e^f\, dx^\mu$, $\hat e^y =
e^h\, dy$, $\hat e^a = e^u\, e^a$, given by
\be
\hat\omega^{\mu y} = f'\, e^{f-h}\, dx^\mu\ ,\qquad
\hat\omega^{ay} = u'\, e^{u-h}\, e^a\ ,\qquad 
\hat\omega^{ab} = \omega^{ab}\ .\label{connection}
\ee
In the present case we have $f=\a\, \varphi+A$, $h=\a\,\varphi -3A$,
and $u=\beta\,\varphi$. Substituting into the $D=11$ supersymmetry 
transformation rule \cite{cjs} 
\be
\delta\psi_M = D_M\epsilon -\ft1{288} (\Gamma_M{}^{N_1\cdots N_4}\, 
F_{N_1\cdots N_4} - 8\, \Gamma^{N_1N_2N_3}\, F_{MN_1N_2N_3})\, 
\epsilon\ ,
\ee
we find that the condition $\delta\psi_\mu=0$ gives
\be
\delta\psi_\mu = \del_\mu\epsilon - \ft12 e^{f-h}\, \Gamma_y\, (
f'\, \Gamma_\mu - \ft16 m\, e^{3\a\varphi -3A}\, \varepsilon_{\mu\nu\rho}\, 
    \Gamma^{\nu\rho})\, \epsilon = 0\ ,
\ee
implying that for preserved supersymmetry we must have
\be
f' = \a\varphi' + A' = -\ft13 m\, e^{3\a\varphi -3A}\ .
\ee
(Actually, we have the freedom here to choose either a plus or a minus
sign on the right-hand side; we choose the minus sign in order to
agree with the choice made previously, where $b_1$ was taken to be
negative.) Similarly, from the condition that $\delta\psi_a=0$, we
obtain another first-order equation, this time for $\varphi$ alone,
namely
\be
\delta\psi_a = D_a\epsilon + \ft12 e^{u-h}\, \Gamma_{ay} (u' - 
\ft16 m\, e^{3\a\varphi-3A}\, \Gamma_{012})\, \epsilon=0\ .
\ee
Note that the explicit indices on $\Gamma_{012}$ here refer to the
vielbein directions in the space of the $x^\mu$ coordinates. From this,
we see that the existence of Killing spinors will require that
\be
u' = -\ft27\a\, \varphi' = \ft16 m\, e^{3\a\varphi -3A} - \kappa\, 
e^{\ft97\a\varphi -3A}\ ,
\ee
since then we will have
\be
\delta\psi_a =(D_a- \ft12\, \kappa\, \Gamma_{ay})\epsilon - \ft1{12} m\, 
e^{\ft{12\a}{7}\varphi}\, \Gamma_{ay}(1-\Gamma_{012})\, \epsilon = 0\ .
\ee
This will have solutions provided that the constant $\kappa$ is given
by $\kappa=\sqrt{\fft{R_7}{42}}$, since then the first expression in
the brackets is the $S^7$-covariant derivative encountered previously
in the Kaluza-Klein reduction of $D=11$ supergravity on the 7-sphere
\cite{dp2}.  In fact, the two first-order conditions that we have just
obtained are precisely the ones given in (\ref{firstorder}) in
appendix \ref{app:b}, where the equations of motion for domain walls are
solved. (Note that we have $b_1 = -\sqrt7\, m/2 <0$ and $b_2
=\sqrt{\ft32 R_7}>0$ here.)  Thus the conditions above reduce to
\bea
\delta\psi_\mu &=& \del_\mu\epsilon + \ft16 m\, e^{3\a\varphi + A}\,
  \Gamma_y
    (\Gamma_\mu +\ft12 \varepsilon_{\mu\nu\rho}\, 
        \Gamma^{\nu\rho})\,\epsilon=0\ ,\nn\\
\delta\psi_a &=& (D_a -\ft12\, \sqrt{\ft{R_7}{42}}\,
                     \Gamma_{ay})\, \epsilon 
    +\ft1{12} m\, e^{\ft{12\a}{7}\varphi}\, 
        \Gamma_{ay}\, (1-\Gamma_{012})\, \epsilon = 0\ ,\label{susy}\\
\delta\psi_y &=& \del_y\epsilon +\ft16 m\, 
       e^{3\a\varphi-3A}\, \Gamma_{012}\, \epsilon=0\ .\nn
\eea
In all of the above expressions, $a$ is a vielbein index in the metric
$ds_7^2$, all indices on Dirac matrices are vielbein indices, the
indices $\mu$ and $y$ on $\psi$ and $\del$ are world indices, and
$\varepsilon_{\mu\nu\rho}$ is the tensor density taking the values
$\pm1, 0$.  It is now easily seen that $\epsilon$ will be a Killing
spinor if it satisfies the conditions
\be
\epsilon= e^{\ft12 f}\, \epsilon_0\otimes \eta = e^{\ft12(\a\varphi+A)}\,
\epsilon_0\otimes\eta\ ,\qquad 
\Gamma_{012}\, \epsilon_0=\epsilon_0\ ,
\ee
where $\epsilon_0$ is a constant spinor in the four-dimensional
spacetime, and $\eta$ is a Killing spinor in the internal
seven-dimensional space, satisfying
\be
D_a\eta -\ft{i}2\, \sqrt{\ft{R_7}{42}}\, \gamma_{a}\, \eta=0
\ .\label{d7ks}
\ee
We are assuming here that the eleven-dimensional Dirac matrices are
decomposed as $\Gamma_\mu=\gamma_\mu\otimes\oneone$, $\Gamma_y =
\gamma_y\otimes \oneone$, and $\Gamma_a=\gamma_5 \otimes \gamma_a$,
with $\gamma_5 = i\, \gamma_{012y}$.  In the case where the internal
space is $S^7$, there will be eight Killing spinors $\eta$ satisfying
(\ref{d7ks}), and so we see that the domain-wall solution, and its
oxidation to $D=11$, preserves one half of the supersymmetry.

     Having seen that the eleven-dimensional metric (\ref{s7d11sol}),
obtained by oxidising the four-dimensional domain-wall solution,
preserves half of the supersymmetry, we now observe that it can in
fact be re-interpreted as the standard BPS membrane solution of $D=11$
supergravity.  To see this, let us introduce a new coordinate $\rho$,
related to the coordinate $y$ in (\ref{s7d11sol}) by
\be
\rho=\sqrt{\fft{42}{R_7}}\, H^{\fft1{15}} = \sqrt{\fft{42}{R_7}}\,
\Big(e^{-\ft5{\sqrt7}\varphi_0} + k y\Big)^{\ft1{15}}\ .\label{rho}
\ee
It is then easy to see that (\ref{s7d11sol}) becomes, upon substituting
(\ref{rho}) for $y>0$ and then continuing the result in $\rho$ so as to
cover the full range $0<\rho<\infty$,  
\be
ds_{11}^2 = \td b_2^{\ft23}\, \Big(1- \fft{\td k}{\rho^6} \Big)^{\ft23}\, 
dx^\mu\, dx_\mu + \Big(1- \fft{\td k}{\rho^6} \Big)^{-2}\, 
d\rho^2 + \rho^2\,d\Omega_7^2\ ,
\ee
where $\td k = \ft16 \, c\, (42/R_7)^{7/2}$, and $d\Omega_7^2$ is the
metric on the unit 7-sphere.  This can be recognised as the standard
form for the $D=11$ BPS membrane solution, written in Schwarzschild
type coordinates.

     At this point, we can make contact with the ``patched'' $D=4$
domain-wall solution mentioned above, which joins together a $k>0$
solution for $y>0$ with a $-k$ solution for $y<0$. Although this seems
unnatural from the $D=4$ viewpoint, what one obtains after oxidation
up to $D=11$ is indeed a natural variant of the $D=11$ membrane
solution. In this solution, $y$ is viewed as a radial coordinate for a
non-isotropic coordinate system, hence naturally bounded to take
values on a half-line. This solution is in fact the maximal analytic
extension \cite{dgt} of the $D=11$ membrane solution \cite{dust}, with
a non-singular horizon at $y=0$ and a central timelike core
singularity at $y=-y_0$.

\subsection{Oxidation of domain walls\label{ssec:2.2}}

    In the previous subsection, we saw that the domain-wall solution
of the $N=8$ four-dimensional supergravity obtained by reducing $D=11$
supergravity on $S^7$ admits a simple interpretation after oxidation
back to $D=11$, namely as the standard membrane solution. Now, we
shall show that in fact this is a rather general feature of the
domain-wall solutions of supergravities reduced on spheres, and in
fact they can all be re-interpreted as extremal $p$-branes back in
their original higher-dimensional theories.

    To see this, we begin by noting that the general result
(\ref{emlag}) for the dimensional reduction of gravity plus an
antisymmetric-tensor field strength of degree $d_x$ or $d_y$ from
$D=d_x+d_y$ dimensions to $d_x$ dimensions is precisely of the form
that we consider in appendix \ref{app:b}, which allowed us to obtain
domain-wall solutions.  Specifically, in the notation of appendix
\ref{app:b}, we have
\be
g_1^2 = m^2\ ,\qquad g_2^2 = 2\, R_y\ ,\qquad a_1= 2\a(d_x-1)\ ,\qquad
a_2 = 2(\a-\b)\ .
\ee
It is easily verified that $a_1$ and $a_2$ are such that the condition
(\ref{lambvan}) is satisfied, implying that indeed we have domain-wall
solutions with $\lambda$ in (\ref{lambdasquare}) vanishing.  In terms
of the parameters $\tilde b_1$, $\tilde b_2$ defined in appendix
\ref{app:b}, the domain-wall solutions are therefore given by
\bea
e^{2\a\varphi} &=& H^{-2d_y/(d_x\, d_y + d_x -2)}\ ,\nn\\
e^{(d_x-1) A} &=& e^{-B} = \td b_1\, H^{(d_x+d_y-2)/(d_x\, d_y + d_x -2)}
+ \td b_2\, H^{(d_x-1) d_y/(d_x\, d_y + d_x -2)} \ .
\eea
Oxidising back to $D$ dimensions, using the formulae given in appendix
\ref{app:a}, we find that the metric becomes
\bea
ds_{\sst D}^2 &=& \Big( \td b_2 + \td b_1 H^{-(d_x-2)(d_y-1)/(d_x\,
d_y + d_x -2)} \Big)^{2/(d_x-1)}\, dx^\mu\, dx_\mu \nn\\
&& +\Big( \td b_2 H^{d_x\, d_y/ (d_x\, d_y + d_x -2)}
+ \td b_1 \, H^{(d_x + 2d_y -2)/(d_x\, d_y + d_x -2)} \Big)^{-2}
\, dy^2\nn\\
&& + \lambda^{-2}\, H^{2(d_x-2)/(d_x\, d_y + d_x -2)}\,  
d\Omega_{d_y}^2 \ ,\label{dwox}
\eea
where $\lambda^2 = R_y/(d_y(d_x-1))$.  

     As in the previous subsection, we may now introduce a new coordinate
$\rho$ in place of $y$, defined here by
\be
\rho = \lambda^{-1}\, H^{(d_x-2)/(d_x\, d_y + d_x -2)}\ .
\ee
In terms of $r$, the oxidised domain-wall metric (\ref{dwox}) becomes
(recalling that $\tilde b_1<0$, $\tilde b_2>0$).
\be
ds_{\sst D}^2 =\td b_2^{2/(d_x-1)}\, 
 \Big(1-\fft{\td k}{\rho^{d_y-1}} \Big)^{2/(d_x -1)}\, dx^\mu\, dx_\mu 
+ \Big(1-\fft{\td k}{\rho^{d_y-1}} \Big)^{-2}\, d\rho^2 + \rho^2\,
d\Omega_{d_y}^2\ ,
\ee
where $\td k = \lambda^{-d_y + 1}\, |\td b_1/\td b_2|$.  Finally, we
make the replacement $x^\mu \rightarrow \td b_2^{-1/(d_x-1)}\, x^\mu$,
and define a new radial coordinate $r$ by $\rho^{d_y-1} = r^{d_y-1} +
\td k$, in terms of which the metric becomes
\be
ds^2 = \Big(1+\fft{\td k}{r^{d_y-1}}\Big)^{-2/(d_x-1)} \, dx^\mu\,
dx_\mu +  \Big(1+\fft{\td k}{r^{d_y-1}}\Big)^{2/(d_y-1)} \,
(dr^2 + r^2\, d\Omega_{d_y}^2)\ .
\ee
This can be recognised as the standard form of an isotropic
non-dilatonic extremal $p$-brane with $p=d_x-2$.  In the context of
supergravity theories, it therefore follows that the domain-wall
solutions that we are considering in this paper all preserve one half
of the supersymmetry. Indeed, in the previous subsection where we
considered the example of the domain wall in $D=4$, we gave an
explicit derivation of this result.

\subsection{Further examples\label{ssec:2.3}}

          It is straightforward to see that the two-parameter
supersymmetric domain-wall solution obtained in section \ref{ssec:2.1}
can be embedded in the Lagrangian (\ref{cp3lag2}) including the
squashing mode, since (\ref{cp3lag2}) reduces to the Lagrangian
(\ref{s7lag}) when the dilaton $\td\phi$ is consistently set to zero.

         The $D=7$ Lagrangian (\ref{d7lag}) fits the pattern of
(\ref{domainlag1}) and (\ref{domainlag2}), with $a_1= 12\a$ and
$a_2=\ft92\a$.  It follows from (\ref{lambdasquare}) that $\lambda=0$.
The two-parameter domain-wall solution is given by
\bea
ds_7^2 &=& e^{2A} dx^\mu dx^\nu\eta_{\mu\nu} + e^{2B} dy^2\ ,\nn\\
e^{-\ft{11}{2\sqrt{10}} \varphi} &=& H=e^{-\ft{11}{2\sqrt10}\varphi_0}
   + k |y|\ ,\qquad B=-6A\ ,\nn\\
e^{6A}&=&\td b_1 H^{\fft3{11}} + \td b_2 H^{\fft8{11}}\ ,
\eea
where $\td b_1= \pm 66m/(5k)$ and $\td b_2=\pm (11/10k)\, \sqrt{3R_4}$.
If we oxidise the solution back to $D=11$, we have
\bea
ds_{11}^2 &=& e^{2\a\varphi + 2A} dx^\mu dx_{\mu} +
        e^{2\a\varphi -12A} dy^2 + e^{2\beta\varphi} ds_4^2(S^4)
\ ,\nn\\
&=&(\td b_1 H^{-\ft5{11}} + 
\td b_2)^{\ft13} \, dx^\mu dx^\nu\eta_{\mu\nu} +
(\td b_1 H^{\fft{13}{33}} + \td b_2 H^{\fft{28}{33}})^{-2}\, dy^2 +
H^{\fft{10}{33}} \, ds_4^2\ ,\nn\\
F_4&=& 6m\, \epsilon_4\ .
\eea
The ${\rm AdS}_7\times S^4$ solution arises in the limit $k\rightarrow
0$ when $\td b_1$ and $\td b_2$ have opposite signs, and
$e^{5\varphi_0/\sqrt{10}} = R_4/(48m^2)$.  When $m=0$, the spacetime
is flat, whilst if $R_4=0$, the solution describes a 5-brane with its
charges uniformly distributed over $ds_4^2$ \cite{lpdomain}.  From the
general discussion in section \ref{ssec:2.2}, we can easily see that
this eleven-dimensional solution is nothing but the standard extremal
5-brane.  This implies in particular that the $D=7$ domain-wall
solution preserves half the supersymmetry.

    Another example of a domain-wall solution arises in $D=5$, in the
theory obtained by reducing type IIB supergravity on $S^5$.  There is
in this case a two-parameter domain-wall solution given by
\bea
ds_5^2 &=& e^{2A} dx^\mu dx^\nu \eta_{\mu\nu} + e^{2B} dy^2\ ,\nn\\
e^{-\ft7{\sqrt{15}}\varphi} &=& H=e^{-\ft7{\sqrt3}\varphi_0} + k|y|
\ ,\qquad B=-4A\ ,\\
e^{4A} &=& \td b_1 H^{\fft27} + \td b_2 H^{\fft57}\ ,\nn
\eea
where $\td b_1=\pm 28m/(3k)$ and $\td b_2=\pm 14/(15k)\, \sqrt{5R_5}$. 
This can be oxidised back to $D=10$, where it becomes
\bea
ds_{10{\rm IIB}}^2 &=&
e^{2\a\varphi + 2A}\, dx^\mu\, dx_\mu + e^{2\a\varphi -8A}\, dy^2
+ e^{2\beta\varphi}\, ds^2(S^5)\ ,\nn\\
&=&(\td b_1 H^{-\ft37} + \td b_2)^{\ft12} dx^\mu dx^\nu\eta_{\mu\nu} +
(\td b_1 H^{\fft{13}{28}} + \td b_2 H^{\fft{25}{28}})^{-2} dy^2 +
H^{\fft3{14}} ds^2(S^5)\ ,\nn\\
H_\5 &=& 4m\, e^{8\a\varphi}\, \epsilon_\5 + 4m\, \epsilon_\5(S^5)\ ,
\nn\\
&=& 4m\, e^{8\a\varphi}\, d^4x\wedge dy + 4m\, \epsilon_\5(S^5)\ .
\label{d5dwox}
\eea
The AdS structure arises when $\td b_1$ and $\td b_2$ have opposite
signs, with $k=0$ and $e^{2\sqrt{\ft{3}{5}}\,\varphi_0} =R_5/(20 m^2)$. 
If $m=0$, the solution is flat spacetime; if $R_5=0$, the solution is a
D3-brane with its charges uniformly distributed on
$ds_5^2$.   From the results given in section \ref{ssec:2.2}, the
metric  (\ref{d5dwox}) can be seen to be equivalent to that of the
standard extremal self-dual 3-brane of type IIB supergravity.  Again, a
consequence is that the $D=5$ domain-wall solution preserves one half
of the supersymmetry.

    Another example of a domain-wall solution arises for the
three-dimensional theory (\ref{d3lag}), obtained by compactifying
six-dimensional supergravity on $S^3$.  It is easily seen that this
admits an AdS$_3$ solution where $\phi$ and $\varphi$ are constants,
given by
\be
e^{\sqrt2\phi_{\sst{\rm AdS}}} = \fft{c}{m}\ ,\qquad
e^{\ft{4\a}{3}\varphi_{\sst{\rm AdS}}} = \fft{2R_3}{3c\, m}\ .
\ee
We may now consistently truncate out the dilaton $\phi$, {\it
i.e.} by setting $\phi=\phi_{\sst{\rm AdS}}$, so that 
the Lagrangian (\ref{d3lag}) then becomes
\be
e^{-1}{\cal L} = R -\ft12 (\del \varphi)^2 -mc\, e^{4\a\varphi} + R_3
e^{\ft{8\a}{3} \varphi}\ ,
\ee
which fits the pattern of the Lagrangian (\ref{domainlag1}) and
(\ref{domainlag2}) with $a_1=4\a$, $a_2=8\a/3$, and $\lambda=0$.  Thus
this Lagrangian admits a domain-wall (\ie string) solution, given by
\bea
ds_3^2 &=& e^{2A} dx^\mu dx_\mu + e^{2B} dy^2\ ,\qquad
e^{-\ft5{\sqrt6}\varphi}=H\equiv e^{-\ft5{\sqrt6}
\varphi_{\sst{\rm AdS}}} + 
k|y|\ ,\nn\\
e^{2A} &=& e^{-B} = \td b_1 H^{\fft25} + \td b_2 H^{\fft35}\ ,
\eea
where $\td b_1 = \pm (5/k)\, \sqrt{mc}$ and $\td b_2 = \pm(5/k)\,
\sqrt{2R_3/3}$.

    Finally, we consider the three-dimensional Lagrangian
(\ref{d3lag2}), obtained by reducing $D=5$ supergravity on $S^2$.  This
Lagrangian fits the pattern of (\ref{domainlag1}) and
(\ref{domainlag2}) with $a_1=4\a$, $a_2=3\a$ and $\lambda=0$, so it
admits a domain-wall solution
\bea
ds_3^2 &=& e^{2A} dx^\mu dx_\mu + e^{2B} dy^2\ ,\qquad
e^{-\ft7{2\sqrt3}\varphi}=H\equiv e^{-\ft7{2\sqrt3}
\varphi_{\sst{\rm AdS}}} + 
k|y|\ ,\nn\\
e^{2A} &=& e^{-B} = \td b_1 H^{\fft37} + \td b_2 H^{\fft47}\ ,
\eea
where $\td b_1 = \pm 7m/(\sqrt3 k) $ and $\td b_2 = \pm(7/k)\,
\sqrt{R_2/2}$.

\section{Cosmological instanton solutions\label{sec:3}}

     The Lagrangians that we obtained in section \ref{sec:1} by
considering the dimensional reductions of supergravities on spheres
have structures that are very similar to those considered in the
context of cosmological instanton solutions in \cite{ht1,ht3}.  In the
simplest cases, for example, where we retain just the metric and the
breathing-mode scalar as dynamical fields, we obtain Lagrangians of
the form ${\cal L} = e\, R -\ft12 e\, (\del\varphi)^2 - e\,
V(\varphi)$.  In this section, we shall examine the cosmological
instanton solutions in the case of the four-dimensional theory
obtained in section \ref{ssec:1.1} by making an $SO(8)$-invariant
truncation of the reduction of $D=11$ supergravity on $S^7$.

     We begin by considering an $SO(4)$-invariant metric ansatz, in the
co-moving frame, given by
\be
ds^2=-d\tau^2 + b(\tau)^2 d\Omega_3^2\ ,\label{d4insmetric}
\ee
where $\tau$ is the co-moving time coordinate and $d\Omega_3^2$ is the
metric on the unit 3-sphere. It follows from (\ref{feq}) and the
formulae in appendix \ref{app:c} that the equations of motion are
given by
\be
\ddot \varphi + 3 \dot \varphi\, \fft{\dot b}{b}=
-V_{,\varphi}(\varphi)\ ,\qquad
\ddot b = -\ft16b\, ({\dot \varphi}^2 - V(\varphi))\ ,
\label{s7eom}
\ee
together with the first-order constraint
\be
\Big(\fft{\dot b}{b}\Big)^2 + \fft{1}{b^2} = \ft1{12} {\dot \varphi}^2 + 
\ft16 V\ ,\label{s7focon}
\ee
where $V$ is the scalar potential, given by (\ref{phipot}).  Note that
the potential has the properties
\bea
V(\varphi_0)&=&0\ ,\qquad {\rm when} \qquad
e^{\ft{24\a}{7} \varphi_0} = \fft{2R_7}{c^2}\ ,\nn\\
V_{,\varphi}(\varphi_0) &=& \fft{24\a}{7}e^{\ft{18\a}{7}\varphi_0} \,
R_7 =\fft{24\a}{7} R_7\, (\fft{2R_7}{c^2})^{\ft34} \, >\, 0
\ .\label{vprop}
\eea
Thus we see that the contribution from $F_4 =c\,e^{6\a\varphi}
\epsilon_4$ plays an important role in allowing the vanishing of the
potential $V$ at a non-singular value $\varphi=\varphi_0$ of the
dilaton.  The potential also has a minimum at 
$\varphi=\varphi_{\sst{\rm AdS}}$: 
\bea 
V_{,\varphi}(\varphi_{\sst{\rm AdS}})
&=& 0\ ,\qquad {\rm when}\qquad e^{\ft{24\a}{7}\varphi_{\sst{\rm
AdS}}} = \fft{6R_7}{7c^2}\ ,\nn\\ 
V_{\rm min} &=&V(\varphi_{\sst{\rm AdS}}) = 
-\ft47 R_7 e^{\ft{18\a}{7}\varphi_{\sst{\rm AdS}}} = 
-\ft47 R_7\, (\fft{6R_7}{7m^2})^{\ft34} \, < \, 0\ .\label{vmin} 
\eea 
Here we are denoting by $\varphi_{\sst{\rm AdS}}$ the value of the
dilaton for which the potential is a minimum.  This is because the
Lagrangian can be further truncated to
\be 
e^{-1} {\cal L} = R - V_{\rm min} 
\ee 
when $\varphi = \varphi_{\sst{\rm AdS}}$, and the equations of motion
following from this Lagrangian allow an AdS solution.  Note that we
have $\varphi_{\sst{\rm AdS}} <\varphi_0$.  Thus we see that this
potential has a shape like a scoop.  At large but negative
$\varphi$. we have $V(\varphi) \rightarrow 0_-$, reducing as $\varphi$
increases, until it reaches $V_{\rm min}$ at
$\varphi=\varphi_{\sst{\rm AdS}}$.  The potential then increases
monotonically to infinity, behaving like a single exponential for
large $\varphi$.

   The equations of motion (\ref{s7eom},\ref{s7focon}) are invariant
under time translations, and hence we can without loss of generality
choose our initial time to be $\tau=0$. Following \cite{ht1}, we
assume that at $\tau=0$, $\dot \varphi(0)=0$ and $V(\varphi(0)) \ge 0$,
and hence $\varphi(0)\ge \varphi_0$.  It then follows from the
constraint (\ref{s7focon}) that at early times the solution for the
metric function $b$ is imaginary \footnote{ One may well ask at this
stage what significance should be attached to imaginary configurations
of fields originally defined to be real. Since, in the quantum theory,
fields are described by Hermitean operators whose eigenvalues are real,
these complex configurations should not be thought of as expectation
values of Hermitian operators but rather as {\it off-diagonal matrix
elements}.  For example, in Minkowski signature a self-dual
(anti-self-dual) Maxwell field is complex and describes the wave
function of a positive (negative) helicity photon i.e.  the matrix
element of the field operator between the vacuum and a one-particle
state of definite helicity \cite{duffisham1,duffisham2}.}
\be
b= i\tau\ ,
\ee
and so the solution is more naturally discussed in a Euclideanised
framework.  In fact in the vicinity of $\tau= 0$ the metric describes
a Euclidean flat space.   

We shall in fact present results for two different Euclidean-signature
theories.  The first, which is the more relevant for the discussion of
cosmological instantons, is obtained by performing a Wick rotation of
the time coordinate in the Minkowski-signature theory.  We shall
discuss this case first, and, afterwards, we shall consider an
alternative Euclidean-signature theory which is obtained by
dimensionally reducing the original $D=11$ theory on AdS$_7$ rather
than $S^7$.

\subsection{The Wick-rotated Euclidean-signature theory\label{ssec:3.1}}

    The question of how Euclideanisation should be performed in
the full supergravity theory is a thorny one, and in the literature
one finds debate on the proper way to Euclideanise antisymmetric
tensor fields, supersymmetry, {\it etc}. In particular, the 
Euclideanisation of the four-form giving rise to the cosmological 
constant has been a subject of much controversy 
\cite{hawking,brown,duff,duncan,ht3}.  
One opinion is that the
electric components of an antisymmetric tensor field should be
imaginary in the Euclidean regime, whereas the magnetic components
should be real \cite{hr}.  
If one wants still to be able to consider electric/magnetic duality
symmetries in the Euclidean regime, then having imaginary electric
charges seems to be unavoidable \cite{hr}.  For example, as observed
in \cite{duffmadore,hr}, although real magnetically-charged extremal
black holes can exist in a Euclideanised theory, electrically-charged
extremal black holes can exist only if the charge, and hence the
electric field, is imaginary. An alternative viewpoint
\cite{duffmadore,duff}, in the original spirit of Belavin, Polyakov,
Schwartz and Tyupkin \cite{bpst} and `t Hooft \cite{thooft} in their
treatment of gauge field instantons, is that all components should be
real so as to ensure that (with the possible exception of gravity
itself) the Euclidean action will be positive semidefinite and bounded
below by the topological charge. This ensures that the Euclidean
functional integrals will be gaussian.

    If we adopt the first of these stances, then in order to Wick
rotate the four-dimensional Lagrangian (\ref{s7lag}) to the Euclidean
regime, we
simply assume that it retains the identical form, but where the metric
tensor is now positive definite.  (In other words, $t\rightarrow -\im
\, \sigma$ is viewed as a general coordinate transformation, with
$\sigma$ subsequently taken to be real.)  The equations of motion will
therefore continue to be written as in (\ref{feq}).  However, there
will be a difference when we dualise the non-propagating 4-form
$F_\4$. Instead of (\ref{f4dual}), we should now solve for $F_\4$ by
writing
\be
F_{\mu\nu\rho\sigma} = \im \, c\, e^{6\a\varphi}\, 
\epsilon_{\mu\nu\rho\sigma}\ ,\label{imag}
\ee
where $c$ is real and $\epsilon_{\mu\nu\rho\sigma}$ is the real
Levi-Civita tensor in the Euclidean regime.  The factor of $\im$ is
necessary since $F_\4$ in four dimensions will necessarily have one
index in the (Wick-rotated) time direction, and so it will necessarily
be electric in character.  Following the standard procedure of
substituting (\ref{imag}) into the remaining equations of motion, we
find that they can be derived from the identical Lagrangian
(\ref{s7lag2}), with potential given by (\ref{phipot}), as we obtained
previously in the Minkowskian regime (with the understanding, as
always, that the metric tensors appearing in (\ref{s7lag2}) are now the
Euclideanised ones). 

     Another piece of ``supporting evidence'' for the appropriateness
of having imaginary electric fields is that by adopting this
prescription, we ensure that the processes of Wick rotation and
dualisation commute: Had we instead insisted that $F_\4$ be dualised
according to (\ref{f4dual}) rather than (\ref{imag}), we would have
found the sign of the $c^2$ term in the potential (\ref{phipot}) to be
reversed.  By contrast, if we simply Wick rotate the already-dualised
Lagrangian given by (\ref{s7lag2}) and (\ref{phipot}), no such sign
reversal occurs.  By requiring that $F_\4$ be imaginary in the
Euclidean regime, we ensure that the order in which the dualisation
and the Wick rotation are performed is immaterial.  One satisfactory
consequence of this is that, regardless of the order of dualisation
and Wick rotation, a solution such as the AdS$_4\times S^7$ solution
in the original Minkowski-signature theory maps over into a sensible
Euclideanised version.

    A further consequence of 
taking the electric components of the field strength
$F_\4$, or, equivalently, the $A_{0ij}$ components of the gauge
potential $A_\3$, 
to be imaginary in the Euclidean regime can be seen by looking at the
Chern-Simons term $F_\4\wedge F_\4 \wedge A_\3$ in the
eleven-dimensional action.  Since the tensor density
$\epsilon^{\mu_1\cdots \mu_{11}}=\pm 1, 0$ is real in both the
Minkowskian and Euclidean regimes, it follows that the Wick rotation
introduces a factor of $\im$ in the Euclideanised Chern-Simons term,
and this becomes consistent with having a real action if the electric
components of $F_\4$, and the $A_{0ij}$ components of $A_\3$, are
imaginary.  It is interesting to note that the various arguments that
can be presented in support of having imaginary electric fields in the
Euclidean regime all seem to revolve around the related notions of
extremality, supersymmetry and duality.

       The metric ansatz for the instanton solution in a 
Euclidean-signatured space takes the form
\be
ds_4^2 = d\sigma^2 + b(\sigma)^2 d\Omega_3^2\ .\label{eucmetric}
\ee
The equations of motion are then given by \cite{ht1} 
\bea
\varphi'' + 3 \varphi' \, \fft{b'}{b} &=&  V_{,\varphi}(\varphi)
\ ,\qquad b'' = -\ft16 b ({\varphi'}^2 + V(\varphi))\ ,
\label{s7eom1}
\eea
together with the first-order constraint
\be
\Big(\fft{b'}{b}\Big)^2  - \fft{1}{b^2} = \ft1{12} {\varphi'}^2 - 
\ft16 V\ .
\label{s7focon1}
\ee
As discussed above, the potential $V(\varphi)$ is given by
(\ref{phipot}) as in the case of a Minkowskian spacetime.  Note that the
equations of motion in the Euclidean-signatured space can be obtained
from those in the Minkowskian-signatured spacetime by making the Wick
rotation $\tau=-i \sigma$.  For both cases, we follow \cite{ht1} and
require that at $\sigma=0$ the potential vanishes, and that in the
vicinity of $\sigma=0$, we have
\be
\varphi(0) \ge \varphi_0\ ,\qquad \varphi'(0) =0\ ,\qquad
b= \sigma\ .\label{initialcon}
\ee

  It follows from (\ref{s7eom1}) that we will have $\varphi''(0_+) =
V_{,\varphi} (\varphi(0_+)) >0$.  This implies that
$(\varphi-\varphi_0)$ is positive, and increases as $\sigma$
increases.  Consequently, $V(\phi)$ is always positive and increases
with $\sigma$, implying that $b''$ is always negative.  Thus the
initial velocity $b'(0) = 1$ reduces with increasing $\sigma$, and
inevitably\footnote{The inevitability is due to the exponential
increase of the potential $V(\varphi)$ at large $\varphi$.  If the
scale size $b$ would expand forever, then it implies that at
$\sigma\rightarrow \infty$, we would have $\varphi\sim \td c \log
\sigma$ where $c$ is a positive constant. Then we find that the
left-hand sides of the equations (\ref{s7eom1}) vanish, while the
right-hand sides diverge at large $\sigma$, for our value of
$V(\varphi)$.} becomes zero at the point where
\be
{b'}^2=b^2(\ft1{12} {\varphi'}^2 +\fft{1}{b^2} -\ft16 V) =0\ .
\label{zerovelocity}
\ee
Since the acceleration $b''$ is still negative, this is a turning
point at which the velocity $b'$ reverses sign and starts to increase,
driving the metric to a singularity at which $b \rightarrow 0$ for
some finite $\sigma_f$.  This evolutionary scenario was discussed in
\cite{ht1}, where analytic continuations were made in order to 
view the solution as describing an open inflationary universe.

    In \cite{ht1}, an assumption was made that the contribution of the
potential $V(\varphi)$ becomes negligible near the singularity, and
thereby it was argued that the solution near the singularity is of the
form
\be
b\sim (\sigma_f -\sigma)^{\ft13}\ ,\qquad \varphi' \sim \fft{2}{\sqrt3}
(\sigma_f -\sigma)^{-1} \label{htsol}\ .
\ee
This solution is independent of the detailed structure of $V(\varphi)$,
and assumes only that it is monotonically increasing, but negligible
near the singularity.  However,
this is not true in our case, where the potential
$V(\varphi)$ is given by (\ref{phipot}), arising from the
$SO(8)$-invariant $S^7$ reduction of eleven-dimensional supergravity,
even though $V(\varphi)$ does increase monotonically in the regime
$(\varphi -\varphi_0)>0$.  To see this, we may substitute the putative
solution (\ref{htsol}) into $V(\varphi)$, to see how it behaves near the
singularity.  We find
\be
V(\varphi) \sim V_{,\varphi}(\varphi) \sim 
(\sigma_f - \sigma)^{-2\sqrt{7/3}}\ .
\ee
This in fact diverges more rapidly than any of the other terms in the
equations of motion (\ref{s7eom1}) and (\ref{s7focon1}), as $\sigma$
approaches $\sigma_f$.  In other words, the HT assumption that the
contribution from $V(\varphi)$ can be neglected near $\sigma=\sigma_f$
is invalid in this case. In fact, in order for the HT solution
(\ref{htsol}) to be valid, the potential $V(\varphi)$, which will be
of order $e^{a\varphi}$ at large $\varphi$, should have $a^2<3$,
and this is not the case for the potential in our Lagrangian.

       To solve the equations with our potential $V(\varphi)$, we note
that for large $\varphi$ we have $V(\varphi) \sim \ft12 c^2\,
e^{6\a\varphi}$.  Assuming that the dilaton diverges logarithmically
at $\sigma_f$, the solution is then given by
\be
b\sim (\sigma_f -\sigma)^{\ft17}\ ,\qquad
e^{-3\a\varphi}\sim \ft74 c\,(\sigma_f - \sigma)\ .\label{newsol}
\ee
(For a generic potential of the form $V(\varphi)\sim \ft12 c^2\,
e^{a\varphi}$ with $a\ge \sqrt3$ at large $\varphi$, we have $b\sim
(r_f -r)^{1/(a^2)}$ and $e^{-a\varphi} \sim a^2
c/(2\sqrt{a^2-3})\,(\sigma_f -\sigma)$.)  Thus we see that the
solution (\ref{newsol}), like the solution (\ref{htsol}), has a
singularity at $\sigma=\sigma_f$, but the degree of the singularity is
different from the one given in \cite{ht1}.

       Note that, owing to the fact that $V(\varphi)$ increases
monotonically in the regime $(\varphi-\varphi_0)>0$, the scalar will
be driven to positive infinity with velocity $\varphi'\ge 0$.  This is
because even if the damping effect resulting from a positive $b'$ were
to tend to reduce $\varphi'$ to zero, it nevertheless would follow
from (\ref{s7eom1}) that the acceleration $\varphi''$ would again be
positive at this point.  Thus, the velocity $\varphi'$ never becomes
negative.  In fact, the potential $V(\varphi)$ monotonically increases
in the regime $\varphi > \varphi_{\sst{\rm AdS}}$. It follows that the
structure of the end-point singularity is independent of the initial
$\varphi(0)$, as long as it is greater than $\varphi_{\sst{\rm AdS}}$.
However, the initial behaviour of the solution does depend on
$\varphi(0)$.  If we choose the initial $\varphi(0)$ such that
$V(\varphi(0))\ge0$, then $b''$ is always negative.  On the other hand, if
we choose $\varphi(0)$ such that $V(\varphi(0))<0$, then the initially
we have $b''>0$, but it will quickly become negative, and will
eventually reverse the expansion of the universe so as to shrink into
the singularity.

         Having obtained the four-dimensional cosmological instanton
solution, it is of interest to examine it from the eleven-dimensional
point of view.  Near the initial time $\sigma\sim 0$, the
eleven-dimensional metric approaches $E^4\times S^7$.  Near the end of
universe, when $\sigma \sim \sigma_f$, the eleven-dimensional metric
is given by
\bea
ds_{11}^2 &=& e^{2\a\varphi}(d\sigma^2 + b^2 d\Omega_3^2) +
e^{2\beta\varphi}ds_7^2\nn\\
&\sim & (\sigma_f -\sigma)^{-2/3}(d\sigma^2 +(\sigma_f-\sigma)^{\ft27}
d\Omega_3^2) + (\sigma_f-\sigma)^{\ft2{21}} ds_7^2\ .
\eea
Thus we see that, owing to the fact that the dilaton diverges towards
$+\infty$ as the final singularity is approached, the size of the
internal seven-sphere $S^7$ shrinks to zero, providing a spontaneous
compactification.  Note that although the size of the three-sphere
$d\Omega_3^2$ shrinks also in the 4-dimensional Einstein metric that we
discussed earlier, its size in the eleven-dimensional metric in fact
tends to infinity.  Defining a co-moving time $\rho\sim (\sigma_f -
\sigma)^{2/3}\rightarrow 0$ in $D=11$, the eleven-dimensional metric
can be written as
\be
ds_{11}^2 \sim d\rho^2 + \rho^{-\ft47} d\Omega_3^2 + \rho^{\ft17} ds_7^2
\ ,\label{d11instantonmet}
\ee from which we can extract a 4-dimensional metric $ds_{\sst{\rm
M}}^2$ using $ds_{11}^2 = ds_{\sst{\rm M}}^2 +
e^{2\beta\varphi} ds_7^2$ (\ie $ds_{\sst{\rm M}}=e^{2\a\varphi}
ds_{\sst{\rm Ein}}^2$), giving
\be
ds_{\sst{\rm M}}^2 \sim d\rho^2 + \rho^{-\ft47} d\Omega_3^2\ ,
\ee
in terms of which the three-sphere expands as $\rho\rightarrow 0$.

        We may also study the instanton solutions in the
four-dimensional theory obtained by reducing $D=11$ supergravity on
the $SU(4)\times U(1)$ invariant squashed 7-sphere.  In this case, the
solution will be supported by all the scalars that are singlets under
$SU(4)\times U(1)$.  It is worth noting that the 3-form field strength
$F_\3$ can be dualised in $D=4$ to give another scalar field.
However, this would be of axionic type, and would come with a
dilatonic prefactor, so it would seem not to fit the bill for the
``inflaton'' in the Hawking-Turok model.  However, let us consider the
equations of motion for an instanton solution.  Take the Lagrangian
(\ref{cp3lag2}), obtained by reducing from $D=11$ on the squashed
$S^7$.  Note that we use the already-rotated dilatons, but we omit the
tildes.  Also, we dualise the 4-form $F_\4$ to a cosmological term
with ``charge" $c$, and we dualise the 3-form $F_\3$ to give an axion
$\chi$.  Thus the Lagrangian in $D=4$ is
\be
e^{-1}\, {\cal L}_4 = R -\ft12 (\del\phi)^2 -\ft12(\del\varphi)^2
 -\ft12 e^{\ft{4}{\sqrt7}\varphi +\ft6{\sqrt{21}} \phi}\, (\del\chi)^2
- \tV(\phi,\varphi)\ ,\label{sqd4lag}
\ee
where the scalar potential $\tV(\phi, \varphi)$ is given by
\be
\tV(\phi,\varphi) = \ft12 c^2\, e^{\sqrt7\varphi}\, 
+6m^2\, e^{\ft{3}{\sqrt7}\varphi +
\ft8{\sqrt{21}}\phi} - e^{\ft3{\sqrt7}\varphi + \ft1{\sqrt{21}}\phi}\,
    R_6\ .\label{sqd4pot}
\ee
The potential has a minimum given by
\bea
\tV_{\rm min} &=& \tV(\varphi_{\sst{\rm AdS}}, \phi_{\sst{\rm AdS}})
=-\ft43 c^2 e^{\ft{4}{\sqrt7}\varphi_{\sst{\rm AdS}}}\ ,\nn\\
e^{\ft7{\sqrt{21}}\phi_{\sst{\rm AdS}}} &=& \fft{R_6}{48m^2}\ ,\qquad
e^{\ft4{\sqrt7}\varphi_{\sst{\rm AdS}}} =
\fft{36m^2}{c^2}(\fft{R_6}{48m^2})^{\ft87}\ .
\eea

     From this we can obtain the equations of motion for the instanton
solution with the metric ansatz (\ref{d4insmetric}); they are given by
\bea
3\fft{\ddot b}{b} &=& -\ft12 {\dot \phi}^2 -\ft12
{\dot\varphi}^2 -
\ft12 e^{\ft{4}{\sqrt7}\varphi +\ft6{\sqrt{21}} \phi}\, {\dot\chi}^2
+\ft12 \tV(\phi,\varphi)\ ,\nn\\
\fft{\ddot b}{b} + 2\Big(\fft{\dot b}{b}\Big)^2 
&=& -\fft2{b^2} +\ft12 \tV(\phi, \varphi)\ ,\nn\\
\ddot \phi + 3 \fft{\dot b}{b}\dot \phi &=&
\fft{3}{\sqrt{21}} e^{\ft{4}{\sqrt7}\varphi +\ft6{\sqrt{21}} \phi}
\, {\dot\chi}^2 - \tV_{,\phi}(\phi,\varphi)\ ,\\
\ddot \varphi + 3 \fft{\dot b}{b} \dot\varphi &=&
\fft{2}{\sqrt7} e^{\ft{4}{\sqrt7}\varphi +\ft6{\sqrt{21}} \phi}
\, {\dot\chi}^2 - \tV_{,\varphi}(\phi, \varphi)\ ,\nn\\
\ddot\chi + 3\fft{\dot b}{b}\dot\chi &=&-\fft{6}{\sqrt{21}} 
\dot \phi\dot\chi 
-\fft{4}{\sqrt7} \dot\varphi \dot\chi\ .\nn
\eea
Note that we can solve the last equation straightforwardly:
\be
\dot \chi = k\, b^{-3}\, 
e^{- \ft{4}{\sqrt7}\varphi -\ft6{\sqrt{21}} \phi}\ .
\ee
Substituting $\dot\chi$ into the above equations, we obtain
\bea
\fft{\ddot b}{b} &=& -\ft16\Big( {\dot \phi}^2 +
{\dot\varphi}^2 + 
k^2 b^{-6} e^{-\ft{4}{\sqrt7}\varphi -\ft6{\sqrt{21}} \phi}
-\ft12 \tV(\phi,\varphi)\Big)\ ,\nn\\
\ddot \phi + 3 \fft{\dot b}{b}\dot \phi &=&
\fft{3k^2}{\sqrt{21}}\,b^{-6}\,e^{-\ft{4}{\sqrt7}\varphi 
-\ft6{\sqrt{21}} \phi}- \tV_{,\phi}(\phi,\varphi)\ ,\\
\ddot \varphi + 3 \fft{\dot b}{b} \dot\varphi &=&
\fft{2k^2}{\sqrt7} \,b^{-6}\, e^{-\ft{4}{\sqrt7}\varphi 
-\ft6{\sqrt{21}} \phi} - \tV_{,\varphi}(\phi, \varphi)\ ,\nn
\eea
together with the first-order constraint
\be
\Big(\fft{\dot b}{b}\Big)^2 + \fft{1}{b^2} =
\ft1{12} {\dot \phi}^2 + \ft1{12}{\dot \varphi}^2 +
\ft1{12} k^2\, b^{-6}\, e^{-\ft{4}{\sqrt7}\varphi 
-\ft6{\sqrt{21}} \phi} + \ft16 \tV(\phi,\varphi)\ .
\ee 

      The contribution of the axion $\chi$ seems to force the solution
to be singular at initial $\tau=0$.  Here we again assume at $\tau=0$,
we have $(\varphi, \phi)=(\varphi_0, \phi_0)$ and $(\dot\varphi,
\dot\phi)=(0,0)$, and $\tV(\varphi_0, \phi_0)=0$.  Assuming that the
universe scale size $b$ is initially infinitesimal, then at small
$\tau$ we will have $b\sim \tau^{1/3}$, which implies a curvature
singularity of the order $\tau^{-2}$.  If we set $k=0$, then the
initial universe will be purely Euclidean, and the subsequent
evolution will be determined by the two-scalar potential
$\tV(\varphi,\phi)$.

         As a final note in this subsection, we may consider solutions
where the unit three-sphere metric $d\Omega_3^2$ in the ansatz
(\ref{eucmetric}) is replaced by a flat metric $d\vec x^2$. In this case,
the equations can be solved exactly, and the solution turns out to be
just the Euclideanisation of the domain wall discussed in section 
\ref{ssec:2.1}. This solution is now given (after a coordinate change
on $y$) by
\bea
ds^2&=& e^{2B} dy^2 + e^{2A} d\vec x^2\ ,\qquad
e^{-\ft5{\sqrt7}\varphi} = e^{-\ft5{\sqrt7}\varphi_{\sst{\rm AdS}}} +
k\, |y|\ ,\nn\\
e^{3A}&=&e^{-B}=\fft{5}{2k}\Big( \sqrt{\ft{6R_7}{7}}\, H^{\fft7{10}} -
c\, H^{\fft3{10}} \Big)\ .\label{flatinstanton1}
\eea
If $k>0$ then $y$ runs from 0, at which the solution is the hyperbolic
space $H_4$, to infinity, at which the solution is flat.  If $k<0$
then $y$ runs from 0, at which the solution is $H_4$, to
$y=-e^{-\ft5{\sqrt7}\varphi_{\sst{\rm AdS}}}/k$, at which one finds a
genuine curvature singularity.  If $k=0$, then the solution is the
hyperbolic space $H_4$ everywhere.  As in our previous discussions of
domain wall in Minkowski-signatured spaces, we can also consider
solutions here where the value of $\varphi$ at $y=0$ differs from the
one chosen in (\ref{flatinstanton1}).  Under these circumstances, the
solution either reaches the $H_4$ form for some value of $|y|$ that is
greater than zero, or else it fails to become $H_4$ by the time $y$
reaches zero.  In either of these cases there is a delta-function
curvature singularity at $y=0$, which can be thought of as a domain wall.

\subsection{Patching cosmological solutions with domain walls
\label{ssec:3.1a}}

         We have seen in the last subsection that one may find
instanton solutions to Wick-rotated supergravity theories that share
some of the features of the HT instanton. One salient feature of such
instantons is the presence of a singularity. In order to avoid such
singularities, a proposal was made in \cite{bous} to patch together
non-singular regions of cosmological instanton solutions using a
domain wall. We shall show that such patching can indeed be carried
out consistently within the specific context of spherically-reduced
$D=11$ supergravity as considered above. The needed domain wall for
this construction has the same source action as that for the
supersymmetric domain-wall solution presented in section
\ref{ssec:2.1}, except that now we shall need to couple it to a
(non-supersymmetric) instanton background (\ref{eucmetric}),
analytically continued back to Minkowski signature.

     In the cosmological analyses of Refs \cite{ht1,bous}, the overall
cosmological solution starts with an Euclidean instanton, which is
followed until a surface of vanishing extrinsic curvature is reached, at
which an analytic continuation back to a Minkowski-signature solution is
made. To do this in our case, one may expand the instanton metric
(\ref{eucmetric}), writing
\be
ds_4^2 = d\sigma^2 + b(\sigma)^2 (d\psi^2 + \sin^2\psi\,d\Omega_2^2)\
.\label{expeucmetric}
\ee
The surface of vanishing extrinsic curvature occurs at $\psi=\pi/2$, at
which one may continue back to a Minkowski-signature metric by letting
$\psi=\pi/2+\im t$. After this analytic continuation, one has the
Minkowski-signature metric
\be
ds_4^2 = -b(\sigma)^2dt^2 + d\sigma^2 + b(\sigma)^2\cosh^2t\,d\Omega_2^2\
,\label{contmetric}
\ee
which describes a universe with de Sitter-like features. The metric
(\ref{contmetric}) describes only part of this spacetime, however, and
needs a further analytic continuation in order to cover the whole
spacetime. Analytic continuation in this way does not provide a refuge
from the singularity, however. The singularity that we have seen in the
last subsection at $\sigma\sim\sigma_f$ persists also in the analytically
continued solution. Accordingly, in \cite{bous} a suggestion was made to
simply cut out the singular portion of the spacetime by patching two
non-singular regions of the form (\ref{contmetric}) together, with a
domain wall located at the patching surface.

     From the viewpoint of $D=11$ supergravity, the coupling of the
needed domain wall to the background supergravity fields is described
by including the action for a fundamental supermembrane \cite{bst}
into the overall (field + source) action. Since we actually wish to
make this coupling in the spherically-reduced $D=4$ theory discussed
in section \ref{sec:1}, we shall choose to let the membrane lie
entirely in the non-compact $D=4$ spacetime, and shall also have to
use the appropriate form of the background metric, as obtained from
the spherical dimensional reduction. The bosonic part of the
supermembrane action for a supermembrane located on the worldvolume
hypersurface $X^{\sst M}(\xi)$ ($\st M=0,1,\ldots,10$) in a background
described by the $D=11$ supergravity fields $g_{\sst{MN}}$,
$A_{\sst{MNP}}$ is
\be
I^{(11)}_{\rm memb} = T\int
d^{3}\xi\,\Big\{ \left[-{\det}\left(\partial_i  X^{\sst M} \partial_j
X^{\sst N} g_{\sst{MN}}(X)\right)\right]^{1/2} +
\ft16\epsilon^{ijk}\partial_iX^{\sst M}\partial_jX^{\sst
N}\partial_kX^{\sst P} A_{\sst{MNP}} \Big\} \ .\label{smembact}
\ee
In order to obtain the corresponding source action in $D=4$
supergravity, we shall adopt the spherical reduction ansatz
(\ref{d11d4red}) and shall restrict the membrane to lie in the $D=4$
subspace. To do this, we divide the $D=11$ coordinates $X^{\sst M}$
into two ranges: $X^{\sst M}=(X^\mu,\ Y^m;\ \mu=0,1,2,3;\
m=4,\ldots,10)$ and then restrict the membrane coordinates by setting
$Y^m={\rm const.}$ The supermembrane action then reduces to
\be
I^{(4)}_{\rm memb} = T\int
d^{3}\xi\, \Big\{ e^{3\alpha\varphi}\left[-{\det}\left(\partial_i  X^\mu
\partial_j X^\nu g^{(4)}_{\mu\nu}(X)\right)\right]^{1/2} +
\ft16\epsilon^{ijk}\partial_iX^\mu\partial_jX^\nu\partial_kX^\rho
A_{\mu\nu\rho}\Big\} \ ,\label{wallact}
\ee
where $g^{(4)}_{\mu\nu}$ is the $D=4$ metric defined in the reduction
ansatz (\ref{d11d4red}) and $\varphi$ is the breathing-mode scalar
field, also defined in (\ref{d11d4red}).

     The consistency of putting a membrane into a given background is
determined by checking whether the ``brane-wave'' equations following
from (\ref{wallact}) upon varying the membrane coordinates $X^{\sst M}$
are satisfied, plus verifying that charge conservation requirements are
satisfied. The brane-wave equations derived from (\ref{wallact}) are
\bea
&&\partial_i\left(e^{3\alpha\varphi}\sqrt{-\tilde g}\tilde
g^{ij}\partial_jX^\nu g^{(4)}_{\rho\nu}(X)\right)
-\ft12e^{3\alpha\varphi}\sqrt{-\tilde g}\tilde
g^{ij}\partial_iX^\mu\partial_jX^\nu
\partial_\rho g^{(4)}_{\mu\nu}(X)\nn\\
&&\hspace{.5cm}-3\alpha\partial_\rho\varphi(X)
e^{3\alpha\varphi}\sqrt{-\tilde
g} - {1\over 3!}\epsilon^{ijk}\partial_iX^\mu\partial_jX^\nu
\partial_kX^\sigma F_{\sigma\mu\nu\rho}(X)\nn\\ &&\hspace{1cm}=0
\,\label{branewave}
\eea
where $\tilde g_{ij}=\partial_iX^\mu\partial_jX^\nu
g^{(4)}_{\mu\nu}(X)$.

     In order to solve the brane-wave equations in the background of
the analytically continued instanton metric (\ref{contmetric}) (with the
4-form field strength now given again by (\ref{f4dual}), since we have
analytically continued back to a Minkowski-signature region), we first
pick an appropriate ``static gauge'' for the membrane worldvolume
coordinates: 
\bea
X^t&=&\xi^0\nn\\
X^\theta&=&\xi^1\nn\\
X^\phi&=&\xi^2\ ,\label{statgauge}
\eea
where $\theta$ and $\phi$ here are the angular coordinates for the $S^2$
part of the metric (\ref{contmetric}).

     We now aim, following \cite{bous}, to patch together two
background solutions of the form (\ref{contmetric}) at an appropriate
value of $\sigma$, which we shall call $\sigma_m$. Accordingly, we shall
try to satisfy the brane-wave equations (\ref{branewave}), together
with the gauge conditions (\ref{statgauge}), by making an  additional
ansatz corresponding to a static spherical membrane at constant $\sigma$:
\be 
X^\sigma=\sigma_m\ .\label{statbraneans}
\ee
The only non-trivial check is that of the brane-wave equation for
$\rho=\sigma$ in (\ref{branewave}), as the other three equations turn out
to be automatically satisfied by virtue of gauge identities conjugate to
the gauge conditions (\ref{statgauge}). Evaluating this one non-trivial
equation, we find the matching condition
\be
{3b'(\sigma_m)\over b(\sigma_m)} =
-3\alpha\varphi'(\sigma_m)e^{3\alpha\varphi(\sigma_m)}
+c\,e^{6\alpha\varphi(\sigma_m)}\ ,\label{matchcond}
\ee
which imposes a relation between the membrane location $\sigma_m$
(\ref{statbraneans}) and the background's integration constant $c$
appearing in (\ref{f4dual}). Note that since the analytically-continued
instanton background (\ref{contmetric}) is not supersymmetric, one
should not have expected a no-force condition for the included membrane
(\ref{wallact}), so it should come as no surprise that there is a
specific consistent value $\sigma_m$ for our static membrane ansatz
(\ref{statbraneans}), dependent on the parameter $c$ determining the
background.

     Finally,  we come to the charge-conservation requirement necessary
for a consistent coupling of the membrane to the background
(\ref{contmetric}). The presence of the source action modifies
the antisymmetric-tensor field equation (\ref{feq}c) by the inclusion of
a delta-function source term:
\be
d(e^{-6\alpha\varphi}\, {*F_\4})=T\, {*J_\3} \ ,\label{sourceFeq}
\ee
where the delta-function source current is 
\be
{*J_\3} = \delta(\sigma-\sigma_m)\, d\sigma\ .
\ee
Clearly, we cannot satisfy this sourced field equation everywhere by the
single condition (\ref{f4dual}) for a given integration constant
$c$, because this single condition would cause the left-hand side of
(\ref{sourceFeq}) to vanish. What we must instead do to achieve a
consistent patching is to have {\em different} values of the integration
constant $c$ on the two sides of the domain wall:
\be
e^{-6\alpha\varphi} \,
{*F_\4}=-\left[c_+\theta(\sigma-\sigma_m)+c_-
\theta(\sigma_m-\sigma)\right]\ .\label{corrsol}
\ee
The charge conservation condition is now transparent: imposing the
sourced antisymmetric-tensor field equation (\ref{sourceFeq}) now yields
the condition
\be
c_- - c_+ = T\ ,\label{chargecons}
\ee
where $T$ is the membrane tension appearing in the source action
(\ref{smembact}, \ref{wallact}). Conditions (\ref{matchcond},
\ref{chargecons}) thus define the $\sigma_m$ location and charge
conservation requirements for a domain wall to be consistently
inserted into an analytically-continued instanton background of the
form (\ref{contmetric}). One should note that at the quantum level,
the tension $T$ of the included membrane will need to be quantised in
accordance with Dirac charge-quantisation conditions taken together
with the web of duality relations for supergravity $p$-branes
\cite{dlm,schwarz,alwis,dirac}. These conditions require the membrane
tension $T$ to sit on a discretised lattice of allowed values. Thus,
the integration constants $c_\pm$ entering into the charge matching
condition (\ref{chargecons}) will have to take values that respect the
lattice of allowed $T$ values.

\subsection{The Euclidean-signature theory from AdS$_7$ reduction
\label{ssec:3.2}}

    Another way of obtaining a Euclidean-signature theory is to
perform a dimensional reduction on a space that includes the time
direction, instead of making a Wick rotation on the time direction.
Although such Euclidean-signature solutions may not be directly
relevant for vacuum tunnelling considerations such as those discussed
in the previous two subsections, we shall nonetheless take the
opportunity to discuss them here. Previous discussions of such
reductions have principally focussed on reductions on
Minkowski-signature tori.  Such compactifications have been considered
in connection with the classification of static $p$-brane solutions in
terms of instanton solutions of non-compact sigma models
\cite{sigma,cllpst}.  Here, we shall instead be concerned with
reduction on an AdS spacetime, as discussed in appendix \ref{app:a}.
Principally, we shall consider the reduction of $D=11$ supergravity on
AdS$_7$.  The generalisation to the other cases, for example $D=11$ on
AdS$_4$ and type IIB on AdS$_5$, is straightforward.

    It is worth emphasising that the Euclidean-signature
supergravities obtained by making timelike reductions from standard
Minkowskian-signature supergravities are themselves perfectly real,
and, by contrast with the Euclidean-signature theories obtained by
Wick rotation, all fields should be taken to be real.  In particular,
the notions of extremality, duality and supersymmetry will now all
require that the fields of the Euclidean-signature theory be real
(see, for example, \cite{cllpst}).

     From the results in appendix \ref{app:a}, we see that after
dualising the 4-form field strength the four-dimensional Lagrangian
will be given by
\be
e^{-1}\, {\cal L} = R -\ft12 (\del\varphi)^2 - \wtd V(\varphi)\ ,
\ee
where the potential $\wtd V(\varphi)$ is given by
\be
\wtd V (\varphi) = -\ft12 c^2\, e^{6\a\varphi} - e^{\ft{18}{7}\a\varphi}
\, R_7\ .\label{wphipot}
\ee
Comparing with the potential $V(\varphi)$ for the usual theory
obtained by dimensional reduction on $S^7$, given in (\ref{phipot}),
we see that the sign of the $c^2$ term here is reversed.  In addition,
although the $R_7$ term is identical in form, it should be recalled
that $R_7$ is negative in (\ref{wphipot}), since it is now the Ricci
scalar of the AdS$_7$ space on which the dimensional reduction is
being performed.  In fact, the net effect is therefore simply to
reverse the sign of the scalar potential that we considered
previously. Thus $\tV$ now has a maximum $\tV_{\rm max} = - V_{\rm
min} >0$. This is true generally: the scalar potential for the
breathing mode in supergravity compactified on an AdS is opposite in
sign to that arising in the compactification on a sphere.

    In this case, we have $\varphi''(0_+) = \wtd
V_{,\varphi}(\varphi_0) <0$.  Thus $(\varphi-\varphi_0)$ in this case
becomes negative as $\sigma$ increases.  The potential $\tV(\varphi)$
is again always positive, implying that the acceleration $b''$ is
always negative.  Thus the velocity $b'$ will decrease. Whether the
end of the universe is singular or not depends on whether the dilaton
stabilises.  We note that the first-order constraint (\ref{s7focon1})
implies in this case that
\be
\Big(\fft{b'}{b}\Big)^2 = \ft1{12} {\varphi'}^2 +
\fft1{b^2} - \ft16 \tV(\varphi)\ .\label{beq}
\ee
Since in the region $(\varphi-\varphi_0)<0$, the potential
$\tV(\varphi)$ is positive, it is possible that the velocity $b'$ can
reach zero at some finite time, and the negative acceleration $b''$
then implies that the universe will eventually shrink down to zero size
at finite $\sigma_f$.  When this happens, there are two possible
outcomes.  The first one is that the dilaton diverges, and runs to
$-\infty$, in which case the potential vanishes.  The solution near
$\sigma_f$ is then given by
\be
b\sim (\sigma_f-\sigma)^{\ft13}\ ,\qquad
\varphi' \sim -\fft{2}{\sqrt3}\, (\sigma_f-\sigma)^{-1}\ ,
\ee
and hence the universe ends up in a singularity.  

        However, there can be a non-singular end-point as well.  Note
that $\tV(\varphi)$ does not increase monotonically for
$(\varphi-\varphi_0)<0$ ; in fact it decreases when $\varphi$ passes
the minimum point $\varphi_{\sst{\rm AdS}}<\varphi_0$.  This implies
that $\varphi''$ can be positive, and hence can cause the value of
$\varphi$ to increase.  The net effect is that $\varphi$ eventually
stabilises at $\varphi_{\sst{\rm AdS}}$.  In this case we
eventually have $b'' = -\ft16 \tV_{\rm max} b$, and so
\be
b\sim \sin(\sqrt{\ft16\tV_{\rm max}}\, (\sigma_f - \sigma))
\ .\label{scenario1}
\ee
In this case, the ``end of the universe'' is an $S^4$. 
The solution has no singularity at $\sigma_f$; in fact the metric
becomes purely Euclidean.  The universe afterwards becomes
periodic.

       {\it A priori}, one would expect that if $\tV_{\rm max}$ were
sufficiently small, such that $b'$ never reached zero, then the
universe would expand forever.  In this case $(\varphi-\varphi_0)$
would be driven to negative infinity, since we have seen above that
the stabilisation of $\varphi$ implies that $b'$ becomes negative
after a certain amount of time.  Whether this possibility arises or
not depends on whether there is such a solution.  It is easy to verify
that the solution exists only for potentials $\tV\sim \ft12 g^2
e^{a\varphi}$ with $a^2 <1/2$.  To see this, we note that
(\ref{s7eom1}) leads to a solution that asymptotically approaches
\be
b\sim \tau^{\ft{1}{a^2}}\ ,\qquad
e^{a\phi} \sim \fft{2\sqrt{3-a^2}}{a^2 g}\, \tau^{-1}\ ,\label{scenario2}
\ee
where $\tau=i\sigma$.  The first-order constraint (\ref{s7focon1})
implies that only if $a^2<1/2$ can the $1/b^2$ contribution can be
ignored. In our case we have $\tV(\varphi)\sim R_7\,
e^{18\a\varphi/7}=R_7\, e^{3\varphi/\sqrt7}$, and hence such a
solution does not occur here.

       Thus we see that there are two possible evolutionary scenarios
in this AdS reduction case, depending presumably on the parameters $c$
and $R_7$ in the potential.  If the slope $\tV_{,\varphi}$ is
sufficiently mild, then $\varphi$ will stabilise at $\varphi_{\sst{\rm
AdS}}$, and the universe will become periodic, without a singularity.
On the other hand if the slope is steeper then $\varphi$ will be
driven to negative infinity, and consequently the universe will end up
at the singular point $\sigma_f$, with a near-singularity solution
given by (\ref{htsol}).

         So far, we have studied the solution in terms of the
four-dimensional Einstein metric $ds_{\sst{\rm Ein}}^2$.  In terms of
the eleven-dimensional metric, the first possibility above, where the
dilaton stabilises to $\varphi=\varphi_{\sst{\rm AdS}}$ and $b$ is
given by (\ref{scenario1}), the size of the internal AdS$_7$
space stabilises also.  The eleven-dimensional metric approaches the form
\be
ds_{11}^2 \sim d\sigma^2 + \sin^2(\sqrt{\ft16\tV_{\rm min}}\, (\sigma_f
-\sigma))\, d\Omega_3^2 + ds_7^2\ 
\ee
when $\sigma \rightarrow \sigma_f$.  In the second possibility, given
by (\ref{scenario2}), the dilaton is instead driven to negative
infinity, and hence the ``internal'' AdS space becomes infinite in
size.  The eleven-dimensional metric is given by
\bea
ds_{11}^2 &\sim& (\sigma_f -\sigma)^{\ft{2\sqrt7}{3\sqrt3}} d\sigma^2 +
(\sigma_f -\sigma)^{\ft23(1+\sqrt{7/3})} d\Omega_3^2 +
(\sigma_f-\sigma)^{-\ft4{3\sqrt{21}}} ds_7^2\nn\\
&\sim& d\rho^2 + \rho^{\ft{2(\sqrt3 +\sqrt7)}{3\sqrt3 +\sqrt7}}
d\Omega_3^2 + \rho^{-\ft{4}{7+3\sqrt{21}}} ds_7^2\ ,
\eea
where $\rho\sim (\sigma_f-\sigma)^{1 + \sqrt{7/27}}$.  Thus from the
eleven-dimensional point of view, this solution really describes an
expanding eight-dimensional universe.

\section{Conclusions}

     In this paper, we have considered various Kaluza-Klein
dimensional reductions of $D=11$ supergravity and type IIB
supergravity, focussing principally on the cases where the internal
space is a sphere.  The goal was to construct lower-dimensional
Lagrangians that could be used for describing cosmological instanton
and domain-wall solutions.  In order that these should also be
solutions of the original $D=11$ or type IIB supergravities, it is
important that the dimensional reduction procedure be a consistent
one.  We achieved this by making a truncation to the subset of
Kaluza-Klein modes that are the singlets under a transitively-acting
symmetry group of the compactifying sphere.  Our main examples were
the $S^7$ and $S^4$ reductions of $D=11$ supergravity, and the $S^5$
reduction of type IIB supergravity, since in all these cases there
exist anti-de Sitter ``vacuum'' solutions in the lower-dimensional
spacetime.  The simplest consistent truncations arise when one just
retains the singlets under the $SO(n+1)$ isometry group of the
$n$-sphere.  In the gravitational sector this corresponds to keeping
only the metric and the breathing-mode scalar in the lower-dimensional
theory.  In the case of spheres of odd dimension, a slightly enlarged
consistent truncation is possible, where one keeps the singlets under
the $SU(m+1)\times U(1)$ subgroup of the $SO(2m+2)$ isometry group of
the ``round'' $(2m+1)$-sphere.  In the gravitational sector this has
the effect of including a further scalar mode in the lower-dimensional
theory, which parameterises the homogeneous ``squashing'' along the
Hopf fibres of $S^{2m+1}$, viewed as a $U(1)$ bundle over $CP^m$.  We
also considered other consistent reductions, on $S^2$, $S^3$ and AdS spaces.

    Having obtained the consistently-truncated lower dimensional
theories, we then studied solutions of two kinds.  One class consisted
of supersymmetric domain-wall solutions, also known as $(D-2)$-branes.
An interesting feature of these solutions is that, once oxidised back
to the original higher-dimensional theory, they become standard isotropic 
$p$-branes.  In fact the domain wall can be understood as arising from
performing a dimensional reduction on the spheres that foliate the
space transverse to the isotropic $p$-brane.  For example, we found a
domain wall ({\it i.e.}\ a membrane) in $D=4$, which can be interpreted
as coming from the dimensional reduction of the eleven-dimensional
M2-brane on the 7-spheres that foliate its transverse space.

    The other class of solutions that we considered consisted of
instanton solutions of the Euclideanised theories, that could, in the
spirit of \cite{ht1,ht3}, be interpreted {\it via} appropriate
analytic continuations as describing open universes. Our
analysis of the instanton solutions concentrated on the four-dimensional
case of greatest phenomenological interest. In our solutions the r\^ole
of the inflaton is played by the breathing-mode scalar of the spherical
reduction, and hence the form of the scalar potential is precisely
determined. This potential is similar to the sorts of potential that
were postulated in \cite{ht1,ht3}, and in particular one finds singular
instanton solutions of the $SO(4)$-invariant deformed $S^4$ kind. 
However, the details of the solutions, and in particular the nature of
the singularity, are somewhat different from the generic ones obtained
in \cite{ht1,ht3}, owing to the fact that certain of their assumptions
about the asymptotic form of the scalar potential are not satisfied by
the explicit potential that we obtained. This is reflected in the fact
that the scale factor $b(\sigma)$ in the instanton metric
(\ref{eucmetric}) near the singularity is of the form $b\sim
(\sigma_f-\sigma)^{1/7}$ rather than of the form $b\sim
(\sigma_f-\sigma)^{1/3}$ which was found in \cite{ht1,ht3}. The precise
origin of the inflaton mode in the 7-sphere compactification of $D=11$
supergravity discussed in \cite{ht3} is not specified.  It is presumably
to be thought of as one of the scalars coming from the Kaluza-Klein
reduction, although from its coupling to the 4-form it is evidently not
the breathing mode.  It is interesting that the outcome from our
analysis where we do use the breathing mode as the inflaton, allowing us
to give an explicit form for its scalar potential, is a singular
instanton solution exhibiting many of the same features that were
obtained in \cite{ht1,ht3} for generic forms of a postulated potential.
The specific potentials that we have obtained in this paper give only a
very small amount of inflation, however. Thus, the search for realistic
inflationary models based on a fundamental underlying theory remains an
open subject.

\newpage
\bigskip\bigskip
\noindent{\large {\bf Acknowledgements and Note Added}}
\bigskip

We are grateful for discussions with Raphael Bousso, Gary Gibbons,
Stephen Hawking, Andr\'e Linde and Neil Turok. We would specifically like
to acknowledge discussions with Rapael Bousso, Stephen Hawking and
Andr\'e Linde that led to clarification of the amount of inflation that
can be expected in the type of model considered in this paper. After
distributing this paper, we received Ref.\ \cite{hrea}, which also
addresses some of the issues discussed in the present work.

\section*{Appendices}
\addcontentsline{toc}{part}{Appendices}        
\appendix

\section{General sphere and AdS reductions\label{app:a}}

Let us consider the general ansatz for a ``breathing-mode'' reduction
on an internal space ${\cal Y}$ which is allowed to be either a
sphere, or an anti-de Sitter spacetime, or, indeed, any other Einstein
space or spacetime, with positive, negative or zero Ricci scalar.  
In the case of reductions on AdS or other spacetimes, the time
coordinate itself forms part of the internal space, and the resulting
lower-dimensional theory is of Euclidean signature.  We shall begin by
considering the reduction of a pure Einstein theory in $D=d_x +d_y$
dimensions, where the lower-dimensional theory has coordinates $x^\mu$
and dimension $d_x$, while the internal space ${\cal Y}$ has
coordinates $y^m$ and dimension $d_y$.  The reduction ansatz for the
$D$-dimensional metric is
\be
d\hat s^2 = e^{2\a\varphi}\, ds_x^2 + e^{2\beta\varphi}\, ds_y^2\ , 
\label{genred}
\ee
where the subscripts $x$ and $y$ indicate the lower-dimensional
space and the compactifying space respectively.  We find
that the reduced action will have a canonical pure Einstein-Hilbert
form for the metric if the constants $\a$ and $\beta$ satisfy
$\a(d_x-2) + \beta\, d_y=0$.  The normalisation of the kinetic term
for $\varphi$ will then be canonical if we choose $\a$ and $\beta$ to be
given by
\be
\a^2 = \fft{d_y}{2(d_x-2)(d_x+d_y-2)}\ ,\qquad \beta = 
-\fft{\a(d_x-2)}{d_y}\ .\label{albet}
\ee
Then we have the following result for the reduction of the
higher-dimensional Einstein-Hilbert Lagrangian:
\be
\hat e\, \hat R = \sqrt{g_y}\, {\cal L}_x\ ,
\ee
where $g_y$ denotes the metric on the $d_y$-dimensional space, and
\be
e^{-1}\, {\cal L}_x = R -\ft12 (\del\varphi)^2 +  
e^{2(\a-\beta)\varphi}\, R_y - \a(d_x-3)\, \square\varphi\ .\label{xyred}
\ee
The last term in (\ref{xyred}) is just a total divergence, and so it
can be dropped.

     For future reference, we record also the expressions for the
vielbein components of the Ricci tensor of the metric $d\hat s^2$,
given in terms of the lower-dimensional quantities for the $x$-space
and $y$-space metrics in (\ref{genred}).  After using the expressions
for $\a$ and $\beta$ given in (\ref{albet}), we find that the vielbein
components of the Ricci tensor are
\bea
\hat R_{\mu\nu} &=& e^{-2\a\varphi}\, \Big(R_{\mu\nu} -\a\,\square\varphi\,
      \eta_{\mu\nu} -\ft12\, \del_\mu\varphi\, \del_\nu\varphi\Big)\ ,\nn\\
\hat R_{mn} &=& e^{-2\beta\varphi}\, R_{mn} -\beta\, e^{-2\a\varphi}\, 
\square\varphi\, \eta_{mn} \ ,\label{ricc}\\
\hat R_{\mu n} &=& 0\ .
\eea 
(The signatures for the local Lorentz metrics $\eta_{\mu\nu}$ and
$\eta_{mn}$ are to be chosen appropriately, depending upon whether a
spacelike or a timelike reduction is being considered.)

   The above discussion may be augmented by including a field strength 
$\hat F_n$ of degree $n$ in the original $D$-dimensional theory, so that
we now start from the Lagrangian
\be
\hat{\cal L} = \hat e\, \hat R - \ft1{2 n!}\, \hat e\, \hat F_n^2\ .
\ee
There are two possible reductions, within the
framework of the consistently-truncated theories that we are considering,
where only singlets under the isometry group of the internal space are
retained.  In the generic case, where the degree $n$ is not equal to
$d_y$, the metric reduction ansatz (\ref{genred}) will be supplemented by
the ansatz
\be
\hat F_n(x,y) = F_n(x)\ .
\ee
On the other hand, in the special case that $n=d_y$, the more general 
ansatz
\be
\hat F_n(x,y) = F_n(x) +  m\, \epsilon_{d_y}\ ,\label{fnred}
\ee
can be made, where $m$ is a constant and $\epsilon_{d_y}$ is the
volume form on the internal space.\footnote{If the internal space
${\cal Y}$ is not a sphere or AdS, it may be that it has additional
invariant tensors aside from the volume form.  In such cases, a
consistent singlet reduction can be performed in which
$\epsilon_{d_y}$ in (\ref{fnred}) is replaced by the invariant tensor.
For instance, if ${\cal Y}$ is a K\"ahler manifold then the K\"ahler
form, or powers of the K\"ahler form, can be used.  Examples of this
kind arise in the squashed $S^7$ and $S^5$ reductions considered in
sections \ref{ssec:1.2} and \ref{ssec:1.5}.}  The
dimensionally-reduced Lagrangian now takes the form
\be
e^{-1}\, 
{\cal L}_x =  R - \ft12 (\del\varphi)^2 + e^{2(\a-\b)\varphi}\, R_y
 -\ft1{2 n!}\, e^{-2(n-1)\a\varphi}\, 
F_n^2 -\ft12 \zeta\, m^2\, e^{2(d_x-1)\a\varphi}\ ,
\ee
where $\zeta=+1$ if the internal space ${\cal Y}$ has Euclidean signature,
and $\zeta=-1$ if ${\cal Y}$ has Minkowski signature.  The last term
is present only if $d_y=n$ and the more general ansatz (\ref{fnred})
is made.

    A case of particular interest arises when $d_x$ is equal to $n$.
We can then dualise the field strength $F_n$, which will have no
degrees of freedom, to give a cosmological-type term.  Specifically,
we will have
\be
F_n = m\, e^{2(n-1)\a\varphi}\, \epsilon_{d_x}\ .
\ee
The two cases
$d_x=n$ or $d_y=n$ can then be treated in a symmetrical way, by taking
$F_n=0$ in the latter case.  The equations of motion in the lower
dimension will then be
\bea
R_{\mu\nu} &=& \ft12 \del_\mu\varphi\, \del_\nu \varphi 
 -\fft1{d_x-2}\, R_y \ e^{2(\a-\b)\varphi}\, g_{\mu\nu} + \fft{\zeta\,
m^2}{2(d_x-2)} \, e^{2(d_x-1)\a\varphi}\, g_{\mu\nu}\ ,\nn\\
\square\varphi &=& -2(\a-\b)\, R_y\, e^{2(\a-\b)\varphi} +
\zeta\, m^2\, (d_x-1)\, \a e^{2(d_x-1)\a\varphi}\ .
\eea
These can be derived from the Lagrangian
\be
e^{-1}\, 
{\cal L}_x =  R - \ft12 (\del\varphi)^2 + e^{2(\a-\b)\varphi}\, R_y
 -\ft12 \zeta\, m^2\, e^{2(d_x-1)\a\varphi}\ .\label{emlag}
\ee

\section{A general class of domain walls\label{app:b}}

     First, consider a general class of Lagrangians in $D$ dimensions 
of the form
\be
e^{-1}\, {\cal L} = R -\ft12(\del\varphi)^2 - V(\varphi)
\label{domainlag1}
\ee
with the potential $V(\varphi)$ given by
\be
V(\varphi) = \ft12 g_1^2\, e^{a_1\varphi} 
-\ft12 g_2^2\, e^{a_2\varphi}\, - \lambda\, g_1\, g_2\, 
e^{\ft12(a_2+a_1)\varphi}\ .
\label{domainlag2}
\ee
If we look for domain-wall solutions where the metric is
\be
ds^2 = e^{2A}\, dx^\mu\, dx_\mu + e^{2B}\, dy^2\ ,\label{gendomainmet}
\ee
where $A$ and $B$ are functions only of $y$, the equations of motion are
\bea
&&\varphi'' + (dA' -B')\varphi' = V_{,\varphi}(\varphi)\ ,\nn\\ 
&&A'' + (dA'-B') A' = dA'' + d(A'-B')A' +\ft12 {\varphi'}^2 =
-\fft{V(\varphi)}{D-2}\ ,
\eea
where we have $d=D-1$.  The equations can be solved by making the ansatz
\bea
\varphi' &=& b_1\, e^{\ft12 a_1\varphi + B} + b_2\, e^{\ft12 a_2\varphi+B}
\ ,\nn\\
A' &=& c_1\, e^{\ft12 a_1\varphi + B} + c_2\, e^{\ft12 a_2\varphi+B}\ ,
\label{firstorder}
\eea
for constants $b_1, b_2, c_1, c_2$, provided that $\lambda$ satisfies
\be
\lambda^2 = -\fft{( (D-2)a_1 a_2 - 2(D-1))^2}{ {[} (D-2) a_1^2 -2(D-1){]}
{[} (D-2)a_2^2 - 2(D -1) {]} }\ .\label{lambdasquare}
\ee
This fist-order ansatz for the solutions is inspired by the equations
that come from requiring supersymmetry in those cases where 
(\ref{domainlag1}) is part of a supergravity Lagrangian.  We have
encountered several examples in this paper where such first-order
equations are indeed seen to arise in this way. 

The dilaton coupling constants $a_1$ and $a_2$ that we consider in this
paper satisfy
\be
a_1 a_2 = \fft{2(D-1)}{D-2}\ ,\label{lambvan}
\ee
and therefore $\lambda=0$.  The domain-wall solutions with non-vanishing
$\lambda$ were constructed in \cite{lpss} for gauged supergravities in
$D=7$ \cite{tn} and $D=6$ \cite{romans}.  Substituting the ansatz to the 
equations of motion, we find that the four constants are given by
\bea
b_1^2 = \fft{(D-2) a_1^2\, g_1^2}{(D-2) a_1^2 -2(D-1)}\ ,&&
c_1 = -\fft{b_1}{(D-2)a_1}\ ,\nn\\
b_2^2 = -\fft{(D-2) a_2^2\, g_2^2}{(D-2)a_2^2 - 2(D-1)}\ ,&&
c_2 = -\fft{b_2}{(D-2)a_2}\ .\label{xygen}
\eea

     Now write $\varphi'' + (dA'-B')\varphi' = \nu_1\, {\varphi'}^2 +
\nu_2\, \varphi'\, A'$.  This is consistent with the above if $\nu_1 =
\ft12(a_2+a_1)$, and $\nu_2 = (D-1) +\ft12(D-2)a_1 a_2 $.  We can
choose $B$ at will by making coordinate transformations of $y$, so we
may choose $B=(d-\nu_2)A= -1/2(D-1)a_1 a_2 A$.  This then implies
$\varphi'' = \nu_1\, {\varphi'}^2$. Thus the domain-wall solution is
given by
\bea
e^{-\ft12(a_1+a_2)\varphi} &=& H\ ,\nn\\
e^{\ft12(D-2)a_1a_2 A} &=& e^{-B} =
\td b_1\, H^{\fft{a_2}{a_2+a_1}} + \td b_2\, H^{\fft{a_1}{a_2+a_1}}
\ ,\label{lambdazerosol}
\eea
where $H$ is the harmonic function $H=e^{-\ft12(a_1+a_2)\varphi_0}+ k\,
|y|$, and $\td b_i = b_i\, (a_1 + a_2) /(2k)$.  If $b_1$ and $b_2$ are
of opposite signs, we can take a limit of $k\rightarrow 0$, in which
the dilaton becomes a constant, given by
\be
e^{(a_1-a_2)\varphi} = e^{(a_1-a_2)\varphi_{\sst{\rm AdS}}}
=- \fft{a_2^2 g_2^2 {[}(D-2)a_1^2 -2(D-1) {]} }{a_1^2 g_1^2
{[} (D-2) a_2^2 -2 (D-1) {]}}\ .
\ee
The metric then becomes AdS$_D$.  This can be easily seen since in
this case, we have $A\sim \log |y|$ and $e^{2B}\sim y^{-2}$.  In all
the examples in our paper, we have $a_1 >a_2$.  Different situations
arise according to whether $\varphi_0$ is chosen to be less than,
greater than, or equal to, $\varphi_{\sst\rm AdS}$.  The solution is
real provided that $b_2>b_1$.  Except when $\varphi_0 =
\varphi_{\sst\rm AdS}$, there is a delta-function curvature
singularity at $y=0$, since the harmonic function is taken to depend
on $|y|$.  One can then think of $y=0$ as defining a domain wall,
separating the two regions of spacetime with $y>0$ and $y<0$.  For
$k>0$, the solution is real for all values of $y$, and the metric is
singular at $|y|=\infty$, where its behaviour is dominated by the
contribution of the $g_2$ cosmological term.  By contrast, for $k<0$
the solution is real for $|y| <y_0\equiv
e^{-\ft12(a_1+a_2)\varphi_0}$, and the metric is singular at $y_0$,
where its behaviour is dominated by the $g_1$ cosmological term.  In
both cases the solution near the horizon at $y=0$ describes two
reflection-symmetric portions of spacetime.  If
$\varphi_0=\varphi_{\sst\rm AdS}$, then the metric becomes the AdS
metric at $y=0$.  If $k>0$ and $\varphi_0>\varphi_{\sst\rm AdS}$, or
$k<0$ and $\varphi_0 <\varphi_{\sst\rm AdS}$, the metrics on the two
sides of $y=0$ reach the AdS form for some value of $|y|$ that is
greater than zero.  On the other hand, if $k>0$ and $\varphi_0
<\varphi_{\sst\rm AdS}$, or $k <0$ and $\varphi_0>\varphi_{\sst\rm
AdS}$, the metrics on the two sides will not have reached the AdS form
when they join at $y=0$.

\section{A general class of cosmological instantons\label{app:c}}

     We consider a general class of Lagrangian in $D$-dimensions of
the form given in (\ref{domainlag1}) and (\ref{domainlag2}) with 
$\lambda=0$, {\it i.e.}\ 
\be
e^{-1} {\cal L} = R -\ft12(\del\varphi)^2 -V(\varphi)\ ,
\label{geninslag}
\ee
where
\be
V(\varphi) =\ft12 g_1^2 e^{a_1\varphi} -\ft12 g_2^2 e^{a_2\varphi}
\label{genpot}
\ee
and dilaton coupling constants $a_1$ and $a_2$ satisfy
\be
a_1 a_2 =\fft{2(D-1)}{D-2}\ .\label{a1a2prod}
\ee
Thus without loss of generality, we can assume that both $a_1$ and
$a_2$ are positive. This scalar potential arises as breathing mode of
spherically reduction of supergravities with admits AdS$\times$Sphere
solution.  For example, M-theory reduction on $S^7$, $S^4$, Type IIB
string on $S^5$ and self-dual three string on $S^3$.  In particular, in
Minkowskian signature these theories all have $a_1 >a_2$.
The potential has the following properties
\bea
V(\varphi_0) &=& 0\ ,\qquad {\rm for} \qquad
e^{(a_1-a_2)\varphi_0} = \fft{g_2^2}{g_1^2}\ ,\nn\\
V_{,\varphi_0} &=& \ft12(a_1-a_2) g_1^{-\ft{2a_2}{a_1-a_2}}
g_2^{\ft{2a_1}{a_1-a_2}}\ ,\label{zeropot}\\
V_{,\varphi}(\varphi_{\sst{\rm AdS}})&=&0\ ,\qquad {\rm for}\qquad
e^{(a_1-a_2)\varphi_{\sst{\rm AdS}}} = \fft{a_2 g_2^2}{a_1 g_2^2}
\ ,\nn\\ 
V_{\rm AdS} = V(\varphi_{\sst{\rm AdS}})& =&
\ft12(a_2-a_1)(g_2^2/a_1)^{\ft{a_1}{a_1-a_2}}
(g_1^2/a_2)^{-\ft{a_2}{a_1 -a_2}}\ .\label{adspot}
\eea
Note that we always have $\varphi_{\sst{\rm AdS}}<\varphi_0$.

       As in the case of $D=4$ discussed in section \ref{ssec:3.1},
the solution is better analysed in Euclidean-signatured space, and
thus the metric ansatz we have for the instanton is of the form 
\be
ds^2 = d\sigma^2 + b(\sigma)^2 d\Omega^2\ .\label{dgenmetric} 
\ee 
The
scalar $\varphi$ also depends only on $\sigma$. The equations of
motion are then given by 
\be \varphi'' + (D-1)\varphi'\,\fft{b'}{b} =
V_{,\varphi}(\varphi) \ ,\qquad b''=-\fft{1}{(D-2)(D-1)}\,
b\,({\varphi'}^2 + V(\varphi))
\label{geneom}
\ee
together with the first-order constraint
\be
\Big(\fft{b'}{b}\Big)^2 -\fft1{b^2}
=\fft{1}{(D-2)(D-1)}\, (\ft12{\varphi'}^2 -V(\varphi))\ .
\label{genfocon}
\ee

       Following \cite{ht1}, we assume at the initial time $\sigma=0$
that $\varphi(0) = \varphi_0$, and hence the potential $V(\varphi)$
vanishes, and also that $\varphi'(0)=0$.  Then at small values of
$\sigma$, the space is purely Euclidean, with $b=v_0 \sigma$, where
$v_0$ is the initial velocity.  The evolutionary structure depends on
whether $a_1>a_2$, or $a_1 <a_2$, corresponding to the two cases
discussed in sections \ref{ssec:3.1} and \ref{ssec:3.2} for $D=4$.

\bigskip\bigskip
{\noindent \underline{Case (1): $a_1>a_2$}}
\bigskip

In this case, from (\ref{zeropot}) we see that 
$V_{,\varphi} (\varphi_0) >0$, so the dilaton field will be driven
towards positive infinity, and it will not stabilise at $\varphi=
\varphi_{\sst{\rm AdS}}$, since $\varphi_{\sst{\rm
AdS}}<\varphi_0$. The negative acceleration $b''$ will eventually
reverse the sign of the velocity, and the universe will end up at a
singularity at finite $\sigma_f$.  At large value of $\varphi$, we
have $V(\varphi)\sim \ft12 g_1^2 e^{a_1\varphi}$, and if $a_1>a_{\rm
crit}$ the potential contribution is less divergent than the other
terms in the equations of motion, and hence the solution becomes
\bea
b\sim (\sigma_f-\sigma)^{\ft{1}{D-1}}\ ,\qquad
\varphi'\sim \fft{D-2}{\sqrt{D-1}}\, (\sigma_f - \sigma)^{-1}\ .
\label{critsol}
\eea
Substituting this solution into the potential, it straightforward to show 
that
\be
a_{\rm crit}^2 = \fft{4(D-1)}{(D-2)^2}\ .\label{acrit}
\ee
Thus for our set of dilaton coupling constants $a_1$, $a_2$,
satisfying (\ref{a1a2prod}), we have $a_1 a_2 >a_{\rm crit}^2$.  (Note
that in $D=4$, we have $a_1 a_2=a_{\rm crit}$.)  Thus in all our
cases, we have $a_1>a_{\rm crit}$.  The late-time solution can then be
obtained by making the following ansatz
\be
b\sim (\sigma_f-\sigma)^{b_1}\ ,\qquad
e^{\ft12a_1\varphi} \sim \fft{b_2}{\sigma_f-\sigma}\ ,
\ee
where the constants $b_1$ and $b_2$ can be determined by substituting this
ansatz into the equations of motion.

\bigskip\bigskip
{\noindent \underline{Case (2): $a_1 <a_2$}}
\bigskip

      In this case, we have $V_{,\varphi}(\varphi_0) <0$, so the
dilaton will be driven towards left.  It will either stabilise at
$\varphi=\varphi_{\sst{\rm AdS}}$, or else $\varphi$ will head towards
$-\infty$, depending on the initial conditions.  Note that
$V(\varphi)$ does not monotonically increase in the region; it has a
maximum value $V_{\rm AdS}$. Two possible scenarios emerge: if the
potential is mild, then the dilaton will stabilise at
$\varphi=\varphi_{\sst{\rm AdS}}$, and the solution will be described
by an AdS Lagrangian $e^{-1} {\cal L} =R - V_{\rm AdS}$, with solution
given by
\be
b\sim \sin(\sqrt{\fft{V_{\rm AdS}}{(D-2)(D-1)}}\, (\sigma_f-\sigma))
\ .
\ee
Note that in this case, the solution has no singularity, and the
universe smoothly shrinks back to zero size, and the metric at
$\sigma\rightarrow \sigma_f$ is again purely Euclidean.  The universe
afterwards become periodic.  Of course, if the sloop of the potential
is sharp, then the dilaton will be driven to infinity, and
the universe will end up at a singularity, with the solution near
$\sigma_f$ of the form (\ref{critsol}).

\section{Curvatures for instanton and cosmological metrics\label{app:d}}

     We look for solutions in $D$ dimensions where the metric has
Euclidean or Minkowskian signature, and takes the form
\bea
\hbox{Euclidean}:&& 
ds^2 = d\sigma^2 + b(\sigma)^2\, d\Omega^2\ ,\label{metricforme}\\
\hbox{Minkowskian}:&& 
ds^2 = -d\tau^2 + b(\tau)^2\, d\Omega^2\ ,\label{metricformm}
\eea
where $d\Omega^2$ is the metric on the unit $n$-sphere.
In the obvious
orthonormal basis $\hat e^0 = d\sigma$ (or $\hat e^0 = d\tau$),
$\hat e^i = b\, e^i$, we find that the vielbein components of the
Riemann tensor are given by
\bea
\hbox{Euclidean}:&& \hat R^i{}_{jk\ell} = \fft1{b^2}\, R^i{}_{jk\ell} -
\Big(\fft{b'}{b}\Big)^2 (\delta^i_{k}\, \delta_{j\ell} - 
\delta^i_{\ell}\, \delta_{jk})
\ ,\nn\\ 
&&\hat R^i{}_{0j0} = -\fft{b''}{b}\, \delta^i_{j}\ \, \\
&&\nn\\
\hbox{Minkowskian}:&& \hat R^i{}_{jk\ell} = \fft{1}{b^2}\, 
R^i{}_{jk\ell} +
\Big(\fft{\dot b}{b}\Big)^2 (\delta^i_{k}\, \delta_{j\ell} - 
\delta^i_{\ell}\, \delta_{jk}) \ ,\nn\\ 
&&\hat R^i{}_{0j0} = -\fft{\ddot b}{b}\, \delta^i_{j}\ . 
\eea
(This is in fact true for an arbitrary metric $d\Omega^2$; it need not
yet be specified to be a unit sphere.)  Thus we find that the vielbein
components of the Ricci tensor are given by
\bea
\hbox{Euclidean}:&&
\hat R_{ij} = \fft{1}{b^2} \Big((n-1)(1-{b'}^2) - b\,b''  \Big)\,
\delta_{ij} \ ,\nn\\
&&\hat R_{00} = -n\, \fft{b''}{b}\ ,\\
&&\nn\\
\hbox{Minkowskian}:&&
\hat R_{ij} = \fft{1}{b^2}\Big( (n-1)(1+ {\dot b}^2) + b\,\ddot b \Big)\,
\delta_{ij} \ ,\nn\\
&&\hat R_{00} = -n \, \fft{\ddot b}{b}\ ,
\eea
where we have now substituted that the metric $d\Omega^2$ is for the
unit $n$-sphere, whose its Ricci tensor is $R_{ij} = (n-1)\,
\delta_{ij}$.   (In fact the Ricci tensor $\hat R_{ab}$ will
take the same form for {\it any} Einstein metric $d\Omega^2$
normalised to have Ricci tensor  $R_{ij} = (n-1)\,
\delta_{ij}$.)

\section{Effective cosmological constants in sphere reductions\label{app:e}}

    Although it is not directly relevant to our calculations in this
paper, it is perhaps worthwhile to investigate the form of the
effective cosmological constant that one obtains in a sphere
reduction, if the breathing mode scalar is eliminated using its
equation of motion.  This result is, however, relevant in the model of
Hawking and Turok \cite{ht3}, which is inspired by the 7-sphere
reduction of $D=11$ supergravity.  They consider a four-dimensional
Lagrangian where the breathing mode does not appear, and has evidently
been integrated out.  (This can be seen from the fact that although
there is an $F_\4^2$ term in their Lagrangian, it is not multiplied by
the exponential of the breathing-mode scalar, unlike the situation in
(\ref{s7lag}).)  The inflaton scalar in the Lagrangian in \cite{ht3}
is presumably, therefore, to be thought of as one of the other scalars
coming from the $S^7$ reduction of $D=11$ supergravity.  One of the
other features of the Lagrangian in \cite{ht3} is that the sign of the
kinetic term for $F_\4^2$ is the opposite of the one that is directly
inherited from $D=11$ under dimensional reduction.  The explanation
for this sign reversal is implicitly contained in \cite{fr}, although
they do not discuss the effective lower-dimensional Lagrangian.  Using
the results obtained in this paper, we can present a more explicit
demonstration of this.

    Consider the four-dimensional Minkowski-signature 
Lagrangian (\ref{s7lag}), which
includes the breathing-mode scalar $\varphi$.  As we have observed,
there exist solutions where $\varphi=\varphi_{\sst AdS}$ is a
constant, which, in terms of the (constant) antisymmetric tensor
$F_\4$ is given by
\be
R_7 = -\ft{7}{144}\, e^{-\fft{60}7 \a\varphi_{\sst\rm AdS}}\, F_\4^2\ ,
\label{adssol}
\ee
as can be seen from (\ref{feq}b).  Substituting this into the equation
of motion (\ref{feq}a) for the metric, one finds 
\bea
R_{\mu\nu} &=& -\ft19 \, e^{-6\a\varphi_{\sst\rm AdS}}\, (F_{\4\mu\nu}^2
-\ft38 F_\4^2\, g_{\mu\nu})\ ,\nn\\
&=& -\ft27\, e^{\fft{18}{7}\a\varphi_{\sst\rm AdS}}\, R_7\, g_{\mu\nu}\ .
\label{neg}
\eea
Thus we see that indeed $R_{\mu\nu}$ is a negative multiple of
$g_{\mu\nu}$, as one expects from the Freund-Rubin AdS$_4\times S^7$
vacuum solution.  

    If one were to derive (\ref{neg}) from a Lagrangian involving only
the metric and some 4-index antisymmetric tensor field strength
$H_\4$, then this Lagrangian would have to have the form
\be
e^{-1}\, {\cal L} = R + \ft1{48}\, H_\4^2\ ,\label{poslag}
\ee
with a positive, rather than negative, sign for the kinetic term for
$H_\4$.  Comparing the equations of motion from this Lagrangian with
(\ref{neg}), we obtain agreement if we choose
\be
H_\4 = \ft{2}{\sqrt3}\, e^{-3\a\varphi_{\sst\rm AdS}}\, F_\4
\ ,\label{hfrel}
\ee
where, however, one still needs to take into account the relation
(\ref{adssol}).

    Although in this sense one can say that the effect of the 7-sphere
compactification of $D=11$ supergravity is to reverse the sign of the
kinetic term for the 4-form field strength, this viewpoint strikes us
as somewhat artificial, since $H_\4$ appearing in (\ref{poslag}) is
not the same as the 4-form field strength $F_\4$ of $D=11$
supergravity.  At the classical level the two are related by
(\ref{hfrel}), but at the quantum level it might be more appropriate
to retain both the original supergravity $F_\4$ and the breathing mode
$\varphi$ in deriving the dynamics of the inflaton field.  If one
wants to take seriously the proposition that the inflationary
mechanism follows from $D=11$ supergravity, then the inflaton
considered in \cite{ht3} must be one of the scalar fields coming from
the $S^7$ reduction.  Even at the classical level, the interactions of
the inflaton will involve the original supergravity 4-form $F_\4$,
the breathing mode $\varphi$, and the scalar curvature $R_7$ of the
7-sphere.  It would therefore be misleading to think of the field
strength $H_\4$ in the effective Lagrangian (\ref{poslag}) as being
synonymous with the true supergravity field strength $F_\4$.

\end{document}